\documentclass[pra,aps,showpacs,twocolumn,amsmath,amssymb,superscriptaddress]{revtex4}

\usepackage{graphicx}
\usepackage{amssymb,amsmath}
\usepackage{color}

\newcommand{\bhat}[4]{\hat{b}_{#1 #2 #3}^{#4}}
\newcommand{\nhat}[3]{\hat{n}_{#1 #2 #3}}

\newcommand{\hc}{\mathrm{h.c.}}
\newcommand{\xvec}{\mathrm{\bf r}}
\newcommand{\avec}{\mathrm{\bf a}}
\newcommand{\aho}{a_{\mathrm{ho}}}
\newcommand{\as}{a_{\rm{s}}}
\newcommand{\Fock}[1]{|#1\rangle_{\rm{F}}}
\newcommand{\FockBra}[1]{{}_{\rm{F}}\langle#1|}
\newcommand{\cik}{c_i^{(k)}}
\newcommand{\cilmk}{c_{i\ell m}^{(k)}}
\newcommand{\njlmi}{n_{j\ell m}^{(i)}}
\newcommand{\Nlm}[1]{N_{\ell m}^{(#1)}}
\newcommand{\nlmi}[1]{n_{#1 \ell m}^{(i)}}
\newcommand{\Klmk}{K_{\ell m}^{(k)}}

\newcommand{\MS}[3]{|\Psi_{#1 #2}^{(\pm)};#3\rangle}
\newcommand{\phiMS}[4]{|\phi_{#1 #2}^{(#3)};#4\rangle}

\begin{document}

\title{Macroscopic Superposition of Ultracold Atoms with Orbital Degrees of Freedom}
\author{M.~A. Garc\'{\i}a-March}
\affiliation{Department of Physics, Colorado School of Mines,
Golden, CO, 80401}
\author{D.~R. Dounas-Frazer}
\affiliation{Department of Physics, University of California,
Berkeley, CA, 94720}
\author{L.~D. Carr}
\affiliation{Department of Physics, Colorado School of Mines,
Golden, CO, 80401}
\date{\today}

\begin{abstract}
We introduce higher dimensions into the problem of Bose-Einstein condensates in a double-well potential, taking into account orbital angular momentum.  We completely characterize the eigenstates of this system, delineating new regimes via both analytical high-order perturbation theory and numerical exact diagonalization.  Among these regimes are mixed Josephson- and Fock-like behavior, crossings in both excited and ground states, and shadows of macroscopic superposition states.
\end{abstract}

\pacs{}

\maketitle

\section{Introduction}\label{sec:Intro}

Bose-Einstein condensates (BECs) in double-well potentials continue to receive much attention due to their wide variety of applications, ranging from precision measurements~\cite{Schumm:2005, Gati:2006,Hall:2007} to optical information processing~\cite{Ginsberg:2007} and quantum computing~\cite{Strauch:2007,Sebby-Strabley:2006,Calarco:2004}. Being a natural realization of a macroscopic two-state system, they provide an ideal system for studying fundamental quantum many-body phenomena and Josephson-type effects~\cite{Legget:2001,Dalfovo:1999}. BECs in double wells exhibit distinct physical regimes.  Three parameters are frequently used to distinguish them: the number of atoms $N$, the tunneling coefficient $J$, and the interaction coefficient $U$.  The two main regimes are called the \textit{Josephson regime} and the \textit{Fock regime.} In the former, tunneling dominates over interactions, $ \zeta/N\gtrsim 1$ with $ \zeta=J/|U|$, and the limiting case is the non-interacting gas, $U=0$. In contrast, in the Fock regime $ \zeta\ll 1$, interactions are much bigger than tunneling, and the limiting case is the infinite-barrier or zero-tunneling case, $J=0$~\cite{Legget:2001,Dalfovo:1999}. These regime designations are based on just two single particle states, one in each well, and most approaches in the literature are an extension of this two-mode concept.  We will show that when more single-particle states are taken into account many new regimes occur.  Moreover, the dimensionality of the double-well potential manifests in the form of a new quantum number, the orbital angular momentum.
\begin{figure}[h]
\begin{center}
\boxed{\includegraphics[width=7.5cm]{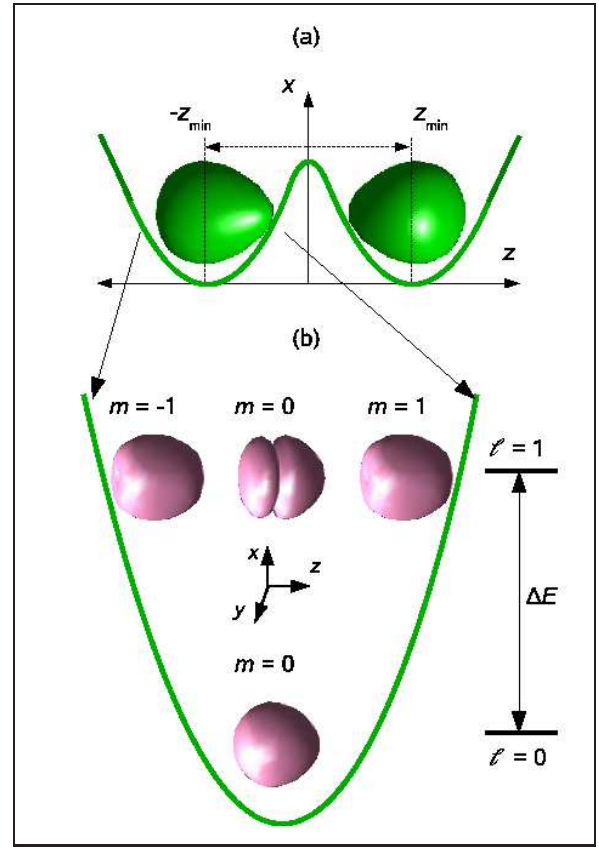}}
\end{center}
\vspace*{-0.5cm}
\caption{(Color online) \textit{Schematic of the potential}. (a) Equipotential surfaces of the double-well potential and (b) zoom of single well and single particle eigenfunctions, where $l$ is the orbital angular momentum and $m$ its $z$ projection.  The first excited level is triply degenerate in 3D. \label{fig:eigenfunctions}}
\end{figure}

In this Article, we use a two-level generalization of the Lipkin-Meshkov-Glick (LMG)  Hamiltonian~\cite{Lipkin:1965,Vidal:2004} to investigate the behavior of ultracold bosons in a \textit{three-dimensional} (3D) double-well potential. In a previous study, we relaxed two assumptions commonly made in similar systems: the symmetric-trap assumption and the one-level assumption~\cite{Dounas-Frazer:2007,Dounas-Frazer:MSthesis}.  We showed that, for reasonable physical parameters, one already requires the first excited state on each side of the double well for the order of 10 to 100 atoms in typical BEC experimental systems.  Therefore, in this study, we consider the effects of this excited state in detail.  The two levels of our 3D Hamiltonian give rise to eight single-particle modes, and atoms in the upper or excited level are allowed to have nonzero orbital angular momentum, as illustrated in Fig.~\ref{fig:eigenfunctions}.  We give  the criteria that permit one to distinguish between Josephson and Fock regimes {\it in both levels}, i.e., we consider also the hopping parameter and interaction parameter for the excited level, leading to novel regimes. We characterize the eigenvectors and eigenvalues in all regimes.  In some regimes different kinds of macroscopic superposition (MS) states, i.e. states for which atoms simultaneosuly occupy both wells, are encountered, ranging from MS states involving atoms only in the bottom level to MS states between atoms only in the excited level with angular momentum, and including {\it mixed} MS states with atoms in both levels. We will illustrate these different kinds of states with surface plots of their probability amplitudes as a function of both energy and their Fock index.  The Fock index orders states in Fock space, as shall be described; the index increases as more and more atoms occupy the upper level. Perturbation theory can give rise to \emph{shadows} which replicate unperturbed patterns in these plots, and appear as faint copies at higher Fock index. We give the criterion to dermacate the regime in which these shadows of MS states occur. 

 Ultracold bosons in double wells are a good candidate for the experimental realization of MS states,  and there are many theoretical proposals in this direction~\cite{Menotti:2001,Higbie:2004,Huang:2006,Piazza:2008,Ferrini:2008,Mazets:2008,Carr:2010, Watanabe:2010}.  One of the main reasons the MS problem has been so heavily pursued, besides technological applications~\cite{Schumm:2005, Gati:2006,Hall:2007,Ginsberg:2007, Strauch:2007,Sebby-Strabley:2006,Calarco:2004}, is because of the possibility of a breakdown in the predictions of quantum mechanics at a macroscopic level~\cite{Legget:2001}. 

In our investigations, we utilize two main methods: numerical exact diagonalization and analytical perturbation theory. We clearly delineate the different regimes in our two-level eight-mode Hamiltonian modeling a 3D double well. Our approach can be extended in a straightforward manner to one- or two-dimensional systems with four or six modes, respectively.  To illustrate the regimes we give numerical examples of every different case using an experimentally realistic double well potential, the Duffing potential. We focus on statics, leaving dynamical considerations, such as tunneling in different regimes, for future work.  Our presentation is organized as follows. In Sec.~\ref{Sec:QTLHam} we describe our model and elucidate the energy scales relevant for the problem.  Once the parameters that determine the different regimes are clearly stated, we characterize the eigenvectors for the different regimes in Sec.~\ref{Sec:CharEig}. In Sec.~\ref{Sec:bounds}, we find criteria for the boundaries between different regimes and for the validity of the one- and two-level approximation. In Sec.~\ref{Sec:Numerics} we illustrate numerically the theoretical results  and find expressions showing how all criteria vary as a function of the number of atoms $N$. In Sec.~\ref{Sec:conclusion} we summarize and discuss our work.  We relegate a detailed description of the construction of single-particle eigenstates and our perturbative methods to appendices.

\section{Quantum Two-Level Approximation}
\label{Sec:QTLHam}

\subsection{Qualitative discussion of physical regimes}

The one-level or two-mode assumption utilized in the original LMG model is valid if coupling to higher single-particle energy levels can be neglected.  Although levels are only completely decoupled when there are no atom-atom interactions, coupling to excited states is very small when the interactions are much weaker than the single-particle energy-level spacing for each well, $\triangle E$.  However, even in this regime effects of the excited level are still important. Eigenvalue crossings, which cannot be described by a one-level approximation, occur when either the number of atoms or the interaction energy is greater than a critical value~\cite{Dounas-Frazer:2007,Dounas-Frazer:MSthesis}.  There are two key crossing regimes in the Fock regime.  First, states with definite occupation of the excited level emerge among the lowest-lying $N+1$ eigenstates. Second, as the interactions are increased further, such crossings involve not only excited states but also the ground state; i.e., when the system is in the ground state one has a finite probability of measuring an atom in an excited level.  In Sec.~\ref{Sec:bounds}, we quantitatively delineate these regimes, and we show that the former crossing does not occur if the condition $N^2|U|\ll \triangle E$ is fulfilled while the latter does not occur if $N|U|\ll \triangle E$. These two key crossing regimes occur also in the Josephson regime for very small interactions, or even vanishing interactions, as detailed below. Finally, we use different criteria to clearly identify the regime of parameters for which the model is valid. In particular, our approach is valid under the diluteness condition, or insofar as the interactions are not too strong.

The very different physical phenomena found in different regimes justify the range of theoretical approaches encountered in the extensive literature on the double-well problem.  Josephson oscillations were first predicted for the limiting case of a non-interacting gas, called the ``extreme Josephson'' or ``Rabi'' regime~\cite{Javanainen:1987}. For non-zero interaction but still in the Josephson regime, mean field (MF) approaches have successfully predicted Josephson effects~\cite{Milburn:1997,Smerzi:1997,Zapata:1998}, while as the atom interaction is increased these models predict macroscopic self-trapping of the condensate in one of the wells~\cite{Milburn:1997,Smerzi:1997,Ostrovskaya:2000,Ananikian:2006,Fu:2006}.  While a number of these works also present the quantum analysis of the system,  a  complete quantum-phase picture of the problem showing the correspondence with the MF approaches, and the transition from delocalized to a fully quantum regime was developed by Mahmud {\it et al}~\cite{Mahmud:2005}. Hence, the MF approach finds its limitation in the Fock regime, and  two-mode approaches find their limitations insofar as more highly excited states are required to describe the dynamics of the problem.  The latter difficulty can be overcome using a multi-mode approach, and thus extending the results to larger interactions~\cite{Ananikian:2006}, but in general, for both cases other methods are required.  One such method is multi-configurational time-dependent Hartree (MCTDH) theory~\cite{Alon:2008}, and its stationary counterpart~\cite{Masiello:2005,Streltsov:2006}.  Other approaches are based on exact diagonalization of the LMG Hamiltonian~\cite{Lipkin:1965,Vidal:2004}, with more than two modes and additional terms~\cite{Dounas-Frazer:2007,Dounas-Frazer:MSthesis, Weiss:2008}. The tunneling dynamics of the system have been studied using MCTDH methods, finding different dynamics than MF methods~\cite{Zollner:2008,Sakmann:2009}. In addition, LMG methods predict exponentially long tunneling times whereas MF methods predict macroscopic self-trapping~\cite{Carr:2010}.  Tunneling and macroscopic self trapping have been observed in experiments~\cite{Shin:2005, Albiez:2005}, though  the latter phenomenon can be attributed to long tunneling times~\cite{Salgueiro:2007,Wang:2006}, small asymmetries in the double-well potential~\cite{Carr:2010}, or even inhomogeneities in the interactions~\cite{Chatterjee:2010}.

Furthermore, the transition from the Josephson regime to the Fock regime, or from a coherent to  an incoherent regime, shows the limitations of the MF approach~\cite{JuliaDiaz:2010}.  The study of this transition elucidated that, in the latter regime, strongly-correlated quantum states appear, showing macroscopic occupation of two single particle states localized at each well~\cite{Pitaevskii:2001,Mahmud:2005,Dounas-Frazer:2007}. These states take the form of MS states, colloquially called Schr\"odinger cat states, and require two independent macroscopic modes that can be coupled or entangled. MS states were also proposed for two-species BECs, i.e., for two internal modes~\cite{Cirac:1998,Ruotekoski:1998,Gordon:1999, Sorensen:2001, Micheli:2003}. A two-species BEC in a single well is mathematically identical~\cite{Steel:1998} to a single-species BEC in a double-well in the one-level approximation.  However, our use of angular momentum makes the double well problem quite different from the usual two-species BEC one.

As atomic interactions are increased, not only MS states appear, but also coupling to states with occupation of the excited levels should be considered, this being responsible for {\it fragmentation} of the condensate~\cite{Spekkens:1999,Mahmud:2005}. Therefore, in this regime, more levels should be included in the model; MCTDH  provides a superior method, but is computationally limited compared to our approach; specifically, to-date MCTDH has not been able to treat 3D systems with orbital angular momentum.  

To make an analogy, helpful for the general reader who may be familiar with the language of ultracold bosons in optical lattices~\cite{Morsch:2006,Lewenstein:2007,Bloch:2008}, suppose we considered not two wells but an infinite number of wells along a line.  This is the lattice problem, and the LMG model becomes the Bose-Hubbard model.  Then our upper level orbital becomes the second band, which is $D$-fold degenerate in $D$-dimensions.  Fragmented states become the Mott-insulator, and the Fock regime is identified with the Mott-insulating regime, while the Josephson regime is identified with the superfluid regime.  However, the Bose-Hubbard model is generally treated for small filling factors of a few atoms per site, while our double-well problem is geared towards large filling factors in hopes of achieving an MS state.

\subsection{General Hamiltonian and double well potential}

The second-quantized Hamiltonian for a system of $N$ interacting bosons of mass $M$ confined by an external potential $V(\xvec)$ at zero temperature is given by
\begin{align}\label{eq:second-quantized}
  \hat{H} =& \int\! d^3\xvec \, \hat{\Psi}^{\dagger}(\xvec)\left[ -\frac{\hbar^2}{2M}\nabla^2 + V(\xvec)\right]\hat{\Psi}(\xvec)\,\nonumber\\
  & + \frac{\bar{g}}{2}\int \! d^3\xvec \, \hat{\Psi}^{\dagger}(\xvec)\hat{\Psi}^{\dagger}(\xvec)\hat{\Psi}(\xvec)\hat{\Psi}(\xvec)\,,
\end{align}
where $\hat{\Psi}(\xvec)$ and $\hat{\Psi}^{\dagger}(\xvec)$ are the bosonic annihilation and creation field operators. The coupling constant $\bar{g}$ depends on the {\it s}-wave scattering
length $\as$ of the atoms, $ \bar{g} = 4\pi\hbar^2\as/M$. 

 We consider a 3D double-well potential with minima at $\xvec = \pm \xvec_{\mathrm{min}}\in\mathbb{R}^3$ and a local maximum at $\xvec = \mathrm{\bf 0}$. Without loss of generality, we will consider a separable potential $V(\xvec) = V_x(x) + V_y(y) + V_z(z)$, built up from two harmonic single-well potentials $V_x(x) + V_y(y)=\frac{1}{2} (\omega_x^2 x^2 + \omega_y^2 y^2$), and a generic 1D double-well potential in the third coordinate $z$,
with two minima at $z=\pm z_{\mathrm{min}}/2$  and a maximum at $z=0$. From here on we call the difference between the maximum and the minima in such a potential the \textit{barrier height}, denoted by $V_0$.
Near a minimum, this 1D potential is   $V(z\pm z_{\mathrm{min}}) \approx \frac{1}{2}\omega^2 z^2$
where $\omega$ is an effective single-well trapping frequency. 

Equation~(\ref{eq:second-quantized}) is valid at low densities, when only binary collisions are relevant, and at low energies, when these collisions are characterized by the {\it s}-wave scattering length of the atoms~\cite{Landau:1977}.  The diluteness condition for a weakly interacting Bose gas is   $\sqrt{|\bar{n}\,a_s^3|} \ll 1$,
where $\bar{n}$ is the average density of the gas.  In the context of the double-well potential, an upper bound on the density of the gas is approximately $\bar{n} = N/(\sqrt{2\pi}\,\aho)^3$, where  $\aho \equiv \sqrt{\hbar/M\omega}$ is the oscillator length and
$\omega$ is the single-well harmonic oscillator frequency.  Correspondingly, we restrict our discussion to the regime
\begin{equation}\label{eq:diluteness}
  N^{1/2} \ll \left|\sqrt{2\pi}\,\aho/\as\right|^{3/2}.
\end{equation}
Although the system is said to be ``weakly interacting'' when condition (\ref{eq:diluteness}) is met, the interaction energy can be on the order of the kinetic energy, or even bigger as corresponds to the Fock regime~\cite{Dalfovo:1999, Legget:2001}.  In Sec.~\ref{Sec:bounds} we obtain condition~(\ref{eq:diluteness}) in terms of the relevant coefficients of the double well problem, thus permitting us to compare this criterion with the criterion characterizing the Fock regime in the numerical results given in Sec.~\ref{Sec:Numerics}.   

\subsection{Two-Level approximation}
\label{sec:two-level}

Double-well potentials in one and two spatial dimensions can be achieved in extremely anisotropic traps. The 1D and 2D transverse trapping frequencies must be sufficiently high to reduce the dimensionality of the single-particle wavefunctions, but should not be near any potential resonances~\cite{Olshanii:1998}. In this Article, we restrict our attention to the 3D case, in particular the axially symmetric one, $ \omega_x=\omega_y$. 

We can expand the field operators in any basis of the Hilbert space.   We use a fixed single-particle basis, constructed from the delocalized eigenfunctions of the single particle Hamiltonian $H_{\mathrm{sp}}=-\frac{\hbar^2}{2M}\nabla^2 + V(\xvec)$.  Our site-localized basis is constructed from appropriate superpositions of delocalized eigenfunctions, analogous to how Wannier states are obtained from Bloch functions on a lattice~\cite{Ashcroft:1976}. This approach results in spatial states of form $ \psi_{n\ell m}(\xvec-\xvec_j)$, where $j$ signifies the left or right well, $n$ is the single-particle energy level, $\ell$ is the orbital angular momentum in 3D, and $m$ is its $z$-projection, as sketched in Fig.~\ref{fig:eigenfunctions}; see also App.~A and Fig.~\ref{fig:eigenfunctionstot} for a more detailed description. Then the field operators can be expanded in this basis as
\begin{equation}\label{eq:hatPsi}
  \hat{\Psi}(\xvec)
  = \sum_{j,n,\ell,m}\bhat{j}{n\ell}{m}{}\psi_{n\ell m}(\xvec-\xvec_j),
\end{equation}
where  $\xvec_{1} \equiv -\xvec_{\mathrm{min}}$ and $\xvec_2 \equiv \xvec_{\mathrm{min}}$ are the minima of the left and right wells. The operators $\bhat{j}{n\ell}{m}{\dagger}$ and $\bhat{j}{n\ell}{m}{}$ satisfy the usual bosonic annihilation and creation commutation relations,
\begin{align}
[\bhat{j}{n\ell}{m}{},\;\bhat{j'}{n'\ell'}{m'}{\dagger}]&=\delta_{jj'}\delta_{nn'}\delta_{\ell\ell'}\delta_{mm'},\nonumber\\
[\bhat{j}{n\ell}{m}{\dagger},\;\bhat{j'}{n'\ell'}{m'}{\dagger}]&=[\bhat{j}{n\ell}{m}{},\;\bhat{j'}{n'\ell'}{m'}{}]=0.
\end{align}

For $\aho\ll z_{\mathrm{min}}$  the functions $\psi_{n\ell m}(\xvec)$ closely resemble the eigenfunctions of the harmonic oscillator potential  $V(\xvec)=\frac{1}{2}\omega^2\xvec^2$:
\begin{equation}\label{eq:psi_nlm}
  \psi_{n\ell m}(\xvec) \approx R_{n\ell}(r)Y_{\ell m}(\theta,\phi),
\end{equation}
for $n \in \{0,1,2,\hdots\}$, $\ell \in \{n,n-2,n-4,\hdots,\ell_{\min}\}$, and $m \in \{-\ell,\;-\ell+1,\;\hdots,\;\ell-1,\;\ell\}$.  Here $R_{n\ell}(r)$ is the radial part of the wavefunction, $Y_{\ell m}(\theta,\phi)$ are the familiar spherical harmonics, and $\ell_{\min}$ is 0 when $n$ is even and 1 when $n$ is odd~\cite{Friedman:1971}.  The energy of an atom associated with the wavefunction $\psi_{n\ell m}(\xvec-\xvec_j)$ is
\begin{equation}\label{eq:E_nlm}
  E_{n} \approx \hbar\omega(n+3/2).
\end{equation}

 We emphasize that the harmonic-oscillator description is only approximate; actual eigenfunctions are distorted from spherical harmonics as sketched in Fig.~\ref{fig:eigenfunctions}.  The two-level approximation, i.e., truncating $n$ at 1, is the lowest order of $n$ at which the dimensionality of the double-well becomes apparent. Because $n = \ell$ for $n\in\{0,1\}$, both the total orbital angular momentum of an atom and its energy level are described by the quantum number $\ell$.  In the two level approximation, the subscript $n$ is superfluous and is hereafter suppressed.

\subsection{Two-Level Hamiltonian}

Substituting Eq.~(\ref{eq:hatPsi}) into the second-quantized Hamiltonian (\ref{eq:second-quantized}) yields the two-level Hamiltonian
\begin{subequations}\label{eq:two-level}
\begin{equation}
  \hat{H} = \hat{H}_{0} + \hat{H}_{1} + \hat{H}_{01},
\end{equation}
where
\begin{align}
  \hat{H}_{\ell} \equiv &\sum_{j,m}\!\!\Bigg\{\!\!
  E_{\ell}\nhat{j}{\ell}{m}
  \!-\!J_{\ell m}\!\!
  \sum_{j'\neq j}\!\Big[ \bhat{j}{\ell}{m}{\dagger}\bhat{j'}{\ell}{m}{} +\hc \Big]
  \nonumber\\
  &+\!\sum_{m'}\!\Big[U^{\ell m}_{\ell m'}
  \nhat{j}{\ell}{m}\left(\nhat{j}{\ell}{m'}\!-\!\delta_{mm'}\right)
  \left(2\!-\!\delta_{mm'}\right)\!\Big]  \nonumber\\
  &+\!\delta_{\ell1}(1-\delta_{m0})U^{10}_{1 1}\!\!
  \left[\!\left(\!\bhat{j}{1}{0}{\dagger}\!\right)^2\!
  \bhat{j}{1}{1}{}\bhat{j}{1}{-1}{}
  +\hc\right]\!\!\!\Bigg\},
  \label{eq:H_1}
\end{align}
and
\begin{align}
  \hat{H}_{01} \equiv \!\sum_{j,m''}\!\!\Bigg\{\!&
      U^{00}_{1m''}\!\!
     \left[\!\left(\!\bhat{j}{0}{0}{\dagger}\!\right)^2\!
     \bhat{j}{1}{m''}{}\bhat{j}{1}{-m''}{}
     \!+\!\hc\right]\nonumber\\
     &\quad\quad\quad\quad\quad\quad
     +4\,U^{00}_{1m''}\,\nhat{j}{0}{0}\,\nhat{j}{1}{m''}\!\Bigg\},
     \label{eq:H_01}
\end{align}
for $j\in\{1,\;2\}$, $\ell\in\{0,\;1\}$, $m,m'\in\{-\ell,\;\hdots,\;\ell\}$ and $m''\in\{-1,0,1\}$. 
Here the number operator $\nhat{j}{\ell}{m}\equiv \bhat{j}{\ell}{m}{\dagger}\bhat{j}{\ell}{m}{}$. The one-level Hamiltonians $\hat{H}_0$ and $\hat{H}_1$ describe atoms in the lowest and first excited energy levels, respectively. Atoms in different energy levels are coupled by the operator $\hat{H}_{01}$. In addition to  the level spacing $\triangle E=E_1-E_0$,  the problem is characterized  by the following  energies,
\begin{equation}
  J_{\ell m}\!=\!-\!\!\!\int\!\!d^3\xvec\,\psi^{\ast}_{\ell m}(\xvec\!-\!\xvec_{\mathrm{min}})\!\!\left[\!-\frac{\hbar^2}{2M}\!\nabla^2\!\!+\!\!V(\xvec)\right]\!\!\psi_{\ell m}(\xvec\!+\!\xvec_{\mathrm{min}}),\label{eq:J_lm}
\end{equation}
and
\begin{equation}\label{eq:U_lm_l'm'}
  U^{\ell m}_{\ell'm'} =
  \frac{\bar{g}}{2}\int \!d^3\xvec\,|\psi_{\ell m}(\xvec)|^2|\psi_{\ell'm'}(\xvec)|^2.
\end{equation}
\end{subequations}

Three basic processes characterize the two-level Hamiltonian: single-atom tunneling between left and right wells, two-atom hopping between energy levels, and atom-atom interactions.  Individual atoms tunnel between wells with energy $J_{\ell m}$. Such transitions do not alter the $z$-component of an atom's angular momentum $m$ as we have chosen $\omega_x=\omega_y$.  Furthermore, single-atom transitions between energy levels $\ell$ are forbidden by the orthogonality of the localized wavefunctions $\psi_{\ell m}(\xvec\pm\xvec_{\mathrm{min}})$; that is, we have chosen a basis such that only interactions can mix levels.   Since one-atom hopping is related to the first term in Eq.~(\ref{eq:second-quantized}), after expressing the field operators in terms of the localized single particle eigenfunctions, all integrals involving pairs of localized functions with different values of $\ell$ and $m$ vanish.    Instead, the second term in Eq.~(\ref{eq:second-quantized}) gives same-site inter-level hopping, which is achieved by \emph{pairs} of atoms.   Notice that  every integral of the form $\int |\psi_{j \ell m}|^2\psi_{j \ell' m'}\psi_{j \ell''m''}$ vanishes, as can be easily shown performing each integral on variable $\theta$ using  the  expressions for the localized functions detailed in App.~\ref{Sec:SP_eigefunctions}.   The only exception is for $\ell'=\ell''=1$ and $m'=1$, $m''=-1$, due to the fact that $\psi_{j \ell-m} = \psi_{j \ell m}^{\ast}$. These give the hopping  terms 
$\left[ \!\left(\!\bhat{j}{1}{0}{\dagger}\!\right)^2\!
  \bhat{j}{1}{1}{}\bhat{j}{1}{-1}{}
  +\hc \right]  $ and $\left[ \!\left(\!\bhat{j}{0}{0}{\dagger}\!\right)^2\!
     \bhat{j}{1}{m''}{}\bhat{j}{1}{-m''}{}
     \!+\!\hc \right]$ that appear in Eq.~(\ref{eq:two-level}).  Then, the only terms that have been neglected  to obtain Eq.~(\ref{eq:second-quantized}) correspond to off-site interactions. These correspond to terms whose integrand is of the form $|\psi_{j \ell m}|^2|\psi_{j' \ell m}|^2$ or 
$\psi^3_{j \ell m}\psi_{j' \ell m}$, which are similar to those considered in~\cite{Spekkens:1999}.
These integrals bring in a term proportional to $\exp\left[-(z_{\mathrm{min}}/\aho)^2\right]$. 
Then, the corresponding interaction coefficients are much smaller than any interaction between atoms in the same well $U^{\ell m}_{\ell'm'}$, which do not show this term, if the barrier height, the distance between wells, or both are not too small.   In particular, the barrier height is big enough to neglect these terms in the Fock-like regimes that capture our interest for MS states.

Thus, according to Eq.~(\ref{eq:second-quantized}), level hopping is induced by atom-atom interactions in the same well, unlike tunneling between wells.  Two interacting atoms hop together between energy levels with energy $U^{00}_{1 m''}$.  A pair of atoms can hop from the lowest to the first excited energy level of the same well in two distinct ways.  Either both atoms enter the $m''=0$ state of the excited level or one atom enters the $m''=+1$ state and the other the $m''=-1$ state.  A similar process, described by the interaction energy $U^{10}_{11}$, occurs within the excited level.  
Pairs of atoms in the same well interact with energy $U^{\ell m}_{\ell'm'}$. 

Here we have used eight modes to expand the field operators in three dimensions. The formalism up to this point is equally valid if we consider more levels, i.e., more modes, yielding to new terms in the Hamiltonian~(\ref{eq:two-level}). On the other hand, to describe 1D and 2D systems, the two-level Hamiltonian (\ref{eq:two-level}) must be modified.  The effects of reducing the number of spatial dimensions are twofold:  the allowed values of the quantum number $m$ become restricted, to $m=0$ for 1D and to $m=\pm 1$ for 2D; and the interaction and tunneling energies are also modified. The 2D problem in fact has less symmetry than the 3D problem, since there is no axis along which the system is invariant under rotations.

\subsection{Characteristic parameters of the double well potential}
\label{Sec:adimensionalization}

As a particular example of a 1D double well potential, we will consider a Duffing  potential  
\begin{equation}\label{eq:Duffing}
 V(z)=\bar{V}_0 \left(-\frac{8}{z_{\mathrm{min}}^2}z^2+\frac{16}{z_{\mathrm{min}}^4}z^4+1 \right),
\end{equation}
with barrier height $\bar{V}_0$.  The Duffing equation arises naturally in MF approaches to the double well problem, where chaotic oscillations of the atomic population at each well are found for time-dependent potentials~\cite{Abdullaev:2000,Lee:2001}; the Duffing potential is also used in elementary portrayals of symmetry-breaking and phase transitions. We use this particular form of the potential to illustrate our results. The results given in Sec.~\ref{Sec:CharEig} are valid for arbitrary double well potentials, since they depend only on the form of the Hamiltonian. Also,  the criteria introduced in Sec.~\ref{Sec:bounds}  hold for arbitrary double well potentials provided that the relevant hopping and interactions coefficients are properly calculated. Instead, in general, it is not possible to represent these criteria for arbitrary potentials in a plane, as we do in Sec.~\ref{Sec:Numerics} for the Duffing potential.  Moreover, this potential gives a straightforward expression for the double well potentials found in experiments and  permits us to characterize the problem using, besides the number of atoms $N$, only two additional parameters.  We show in the following that these two parameters are related to the  barrier height $\bar{V}_0$, the distance between minima $z_{\mathrm{min}}$, and the coupling constant $\bar{g}$.

For the Duffing potential~(\ref{eq:Duffing})  $\omega=32 \bar{V}_0/z_{\mathrm{min}}^2$ and, for $\omega_x=\omega_y=\omega$, we have $V_x(x)=16 \bar{V}_{0}\,x^{2}/z_{\mathrm{min}}^{2}$ and likewise for $V_y$. For an atom species of mass $M$, we numerically calculate the eigenfunctions $ \psi_{n\ell m}(\xvec-\xvec_j)$, the energy levels $E_{\ell}$,  and the hopping coefficients~(\ref{eq:J_lm}). Finally,    for a number of atoms $N$, we completely characterize the system once $\bar{g}$ is known and the coefficients~(\ref{eq:U_lm_l'm'}) are found. As we will see,  we can distinguish different regimes in terms of criteria that relate the level spacing, the number of atoms, and the hopping and interaction coefficients. In this section, we describe a procedure to  express these coefficients in such a manner as to clearly identify these regimes for any atomic species.

The recoil energy associated with a 1D periodic optical lattice of wavelength $\lambda$ is defined as $E_r=2\hbar^2 \pi^2/M\lambda^2$. Analogously, for the 1D Duffing potential, we consider $\lambda=2 z_{\mathrm{min}}$ and $E_r=\hbar^2 \pi^2/2 z_{\mathrm{min}}^2 M$. Dividing the potential by $E_r$ we can write $\tilde{V}(z)=V_0\left(-8 z^2/z_{\mathrm{min}}^2+16 z^4/z_{\mathrm{min}}^4+1\right)$ with 
\begin{equation}\label{eq:V0}
V_0=\frac{2M \bar{V}_0 z_{\mathrm{min}}^2}{\hbar^2 \pi^2}. 
\end{equation}
Similarly, $\tilde{V}_x(x)=16 V_0 x^{2}/z_{\mathrm{min}}^2$, and likewise for $\tilde{V}_y(y)$.
Then,
\[
\frac{J_{\ell m}}{E_{r}}=\int d\xvec\tilde{\Phi}_{j \ell m}(\tilde{\xvec})^{*}\left(-\frac{1}{\pi^{2}}\nabla^{2}+\tilde{V}(\tilde{\xvec})\right)\tilde{\Phi}_{j' \ell m}(\tilde{\xvec}),\]
where $\tilde{V}(\tilde{\xvec})\equiv\tilde{V}_x(\tilde{x})+\tilde{V}_y(\tilde{y})+\tilde{V}_z(\tilde{z})$,
$(\tilde{x},\tilde{y},\tilde{z})=(1/z_{\mathrm{min}})(x,y,z)$, and
$\tilde{\Phi}_{j \ell m}(\tilde{\xvec})=z_{\mathrm{min}}^{3/2}\psi_{\ell m}(\tilde{\xvec}\pm\tilde{\avec})$.
Also, \[
\frac{U^{\ell m}_{\ell' m'}}{E_r}=g\int d\tilde{\xvec}|\tilde{\Phi}_{j \ell m}(\tilde{\xvec})|^{2}|\tilde{\Phi}_{j \ell'm'}(\tilde{\xvec})|^{2},\]
with 
\begin{equation}\label{eq:g0}
g=\frac{\bar{g}}{2 E_{r}z_{\mathrm{min}}^{3}}.
\end{equation}
 Finally, the functions $\tilde{\Phi}_{j \ell m}(\tilde{\xvec})$ and their corresponding eigenvalues can be numerically calculated as detailed in App.~\ref{Sec:SP_eigefunctions}.  Therefore, given the number of atoms $N$, we can characterize the problem with two
parameters, i.e., $V_{0}$ and
$g$.  Notice that the distance between wells $z_{\mathrm{min}}$ is enclosed in the scaling procedure.

\subsection{Energy scales in the high and low barrier limit}

Let us now obtain, in the high barrier limit, the relationships between the different relevant energy scales of the problem, i.e., the level spacing $\triangle E$,  $ U^{\ell m}_{\ell' m'}$, and $ J_{\ell m}$.  In App.~\ref{Sec:Exprs_Coef} it is shown, approximating the eigenfunctions with the spherical harmonics, that the interaction energy of two atoms occupying the lowest energy level of one well is
\begin{equation}\label{eq:U0}
  U_0\equiv U^{00}_{00}= g\left(\frac{\alpha}{2\pi}\right)^{\frac{3}{2}},
\end{equation}
where $\alpha\equiv4\pi\sqrt{V_0}$.  Also, the following relationships can be obtained among the hopping and interaction energies,  in the high barrier limit:
\begin{equation}
\label{Eq:Uscales}
  U^{00}_{10}  = U^{00}_{1\pm1} =  U^{1\pm1}_{1\pm1} = 2U^{10}_{1\pm1}  = (2/3)U^{10}_{10}= U_0/2.
\end{equation}

Finally, it is also shown in App.~\ref{Sec:SP_eigefunctions}  that the tunneling energies satisfy
\begin{equation}
\label{Eq:Jscales}
J_{00} =(3/4) J_{1\pm1},\quad J_{10}\gg J_{00}\,,
\end{equation}
provided that $\triangle E$ is not small. For notational simplicity we define
\begin{equation}
  J_0 \equiv J_{00},\quad  \mathrm{and} \quad J_1 \equiv J_{10}.
\end{equation}

Thus, in this high-barrier regime, there are four relevant energies: the energy level spacing, $\triangle E$; the tunneling energies, $J_0$ and $J_1$; and the interaction energy, $U_0$. Nevertheless, all these energy scales are not independent parameters since they depend solely on $V_{0}$ and $g$.

 These results hold as far as the spherical harmonics are a good approximation of the eigenfunctions of the double well, i.e., for high $V_{0}$. For the low barrier limit, we use the expressions given in App.~\ref{Sec:Exprs_Coef}, with numerical evaluation of the corresponding eigenfunctions and eigenvalues. In both limits, once the  parameters, $V_0$, and $g$ are given, all coefficients can be calculated, and, for a given number of atoms $N$, all many-body eigenstates can be found. Thus the Duffing potential makes this a three parameter problem.  Although we restrict our discussion to repulsive interactions, $U_0>0$,  our results hold for $U_0 < 0$ as well.

\subsection{Fock basis and dimension of Hilbert space}
Throughout our discussion, we operate in Fock space. An arbitrary state vector $|\Psi\rangle$ in Fock space has the following representation,
\begin{equation}\label{eq:PsiFock}
  |\Psi\rangle = \sum_{i=0}^{\Omega - 1}c_i\Fock{i},\quad
  \Fock{i} = \bigotimes_{j,\ell,m}|\njlmi\rangle,
\end{equation}
where
\begin{equation}
  |\njlmi\rangle = \frac{1}{\sqrt{\njlmi!}}\left(\bhat{j}{\ell}{m}{\dagger}\right)^{\njlmi}|0\rangle.
\end{equation}
Here $\Omega$ is the dimension of the Hilbert space $\{\Fock{i}\}$, $i$ is the Fock-space index, and $|c_i|^2$ is the probability of finding $\njlmi$ atoms in the $\ell$th energy level of the $j$th well with $z$-component of angular momentum $m$ when the system is described by state $|\Psi\rangle$.  We work in the canonical ensemble, i.e., we require the total number of atoms
\begin{equation}\label{eq:Ntotal}
  N = \sum_{j\ell m}\njlmi,
\end{equation}
to be constant.  Under this restriction, the dimension of the Hilbert space $\Fock{i}$  is given by
\begin{equation}
  \Omega = \frac{(N+n_m-1)!}{N!\;(n_m-1)!},
\end{equation}
where $n_m$ is the number of modes used to expand the field operator. For the double well in 3D we have $\ell$ truncated at 1, so $n_m=8$.  For a large number of atoms, $\Omega$ scales like $N^7$. 

The index $i$ is chosen to increase with the number of atoms in well $j=1$ of the  lower level,  with the number of atoms in the same well in the excited level with $m=-1$, then with $m=0$, and finally  with $m=1$. Therefore,  for the first $N+1$ Fock vectors $i=1+n_{100}$ and they correspond to vectors with no occupation of the excited level. Thus, they satisfy
\begin{equation}
  \Fock{i} =  \frac{1}{\sqrt{n_{100}!n_{200}!}}\!\left(\!\bhat{1}{0}{0}{\dagger}\!\right)^{n_{100}}\!\left(\!\bhat{2}{0}{0}{\dagger}\!\right)^{n_{200}}|0\rangle,
\end{equation}
for $i =0,\;1,\;\hdots,\;N$.   The one-level approximation can easily be recovered from Equation (\ref{eq:two-level}) by requiring $i\leq N+1$.  In this truncated space, the dimension of the Hilbert space reduces to that of the one-level approximation, namely, $N+1$,  the two-level Hamiltonian $\hat{H}$ reduces to the one-level Hamiltonian $\hat{H}_{0}$, and we recover the LMG Hamiltonian. 

For the next $6N$ vectors,  one atom occupies the excited level and
\begin{align}
  i=& 2+n_{100}+N\left[\sum_m n_{11m}+(2N_{10}+4N_{11}+1)\right]\nonumber
\end{align}
where $N_{\ell m}$ is the number of atoms at level $\ell$ with z-component of the angular momentum $m$. These vectors  correspond to all combinations of $N-1$ atoms in the lower level and a single atom in the excited level with  $m=-1,\,0,\,1$, in two wells. The Fock index $i$ increases further  with all combinations of $p=2,\hdots,N$ atoms occupying the excited levels and $N-p$ atoms in the lower level.

\section{Characterization of Eigenstates}
\label{Sec:CharEig}

We begin our analysis with a characterization of the eigenstates $|\phi_k\rangle$ of the two-level Hamiltonian (\ref{eq:two-level}).  The eigenstates satisfy
\begin{equation}
  \hat{H}|\phi_k\rangle = \varepsilon_k|\phi_k\rangle,
\end{equation}
where $\varepsilon_k$ is the energy eigenvalue corresponding to the state $|\phi_k\rangle$.  The eigenstate label $k$ is chosen to increase with $\varepsilon_k$.  In order to describe these states, we will use the Fock-space amplitudes
\begin{equation}
  \cik = \FockBra{i}\phi_k\rangle.
\end{equation}

Insofar as the interlevel effects are not relevant, the eigenstates fall into one of two categories: \emph{harmonic-oscillator-like states} (HO states) or MS states. When the barrier between wells is low, $\zeta/N\gtrsim 1$, all states are harmonic oscillator-like. This regime is known as the {\it Josephson regime}. On the other hand, MS states dominate the spectrum in the high barrier limit, $\zeta\ll 1$. This limit is known as the {\it Fock regime}. We recall that our coefficients $ U_{\ell m}^{\ell' m'}$ and hopping coefficients, $ J_{\ell}$ depend on  $\ell$ and $m$. Indeed, the tunneling coefficient for the excited atoms with $m=0$, $J_1$, is much bigger than  $J_0$. Then, between these two regimes, an mixed one can be found, for which MS states occur for atoms in the bottom level while HO states occur for the excited ones.  When the level spacing is comparable to $N^2U_{0}$, interlevel effects can be no longer neglected, and another category of eigenstates emerges. These states show coupling between MS states with atoms only in the lowest energy level and states with atoms in the excited one;  we name them \emph{shadows of the MS states}.

\subsection{Non-interacting limit: harmonic-oscillator like states}

We first consider the limit $\zeta/N\rightarrow\infty$ by setting $NU_0 = 0$, i.e., the non-interacting limit.  In this case, both the energy levels and the orbital states are completely decoupled.  The two-level Hamiltonian Eq.~(\ref{eq:two-level}) is thus reducible to four independent \emph{one-level} Hamiltonians, since all couplings between different  $\ell$ and $m$ depend on interactions.  Furthermore, because $[\hat{H},\nhat{1}{\ell}{m}+\nhat{2}{\ell}{m}] = 0$ for all $\ell$ and $m$, the eigenstates of the two-level Hamiltonian must have definite occupation of the $m$th orbital state of the $\ell$th energy level. Let $\Nlm{k}$ be the number of atoms occupying the $m$th orbital state of the $\ell$th energy level for the $k$th eigenstate. Then, $\sum_{\ell,m}\Nlm{k} = N$. Let us denote the one-level eigenstates as $|\phi_{\Klmk}\rangle$, where $\Klmk=0,1,\dots,\Nlm{k}$ is the one-level eigenstate label. Then, the $k$th eigenstate $|\phi_k\rangle$ is a direct product of these one-level eigenstates:
\begin{equation}\label{eq:HO_states}
  |\phi_k\rangle = \bigotimes_{\ell,m}|\phi_{\Klmk}\rangle.
\end{equation}
Likewise, the $i$th Fock space amplitude $\cik$ and the eigenenergy $\varepsilon_k$ can be expressed in terms of the one-level amplitudes and energies as
\begin{equation}\label{eq:cik}
  \cik = \prod_{\ell,m}\cilmk\,\delta_{\Nlm{i}\Nlm{k}},
\end{equation}
and
\begin{equation}\label{eq:eigenval}
  \varepsilon_k = \sum_{\ell,m}\varepsilon_{k\ell m},
\end{equation}
where $\Nlm{i} \equiv \nlmi{1} + \nlmi{2}$. Here, $\cilmk$ is the one-level amplitudes  and energies $\varepsilon_{\ell m k}$  are the one-level energies.  Both quantities can be obtained exactly in the non-interacting limit. The amplitudes are given by
\begin{equation}\label{eq:cilmlcoeff}
  \cilmk \!\!= \! a\!_{\Klmk}\!\!\left(\!\Nlm{k}\!\right)\!
  h\!_{\Klmk}\!\!\left(\!\nlmi{1}\!\left|\Nlm{k}\right.\!\right)\!
  p\!\left(\!\nlmi{1}\!\left|\Nlm{k}\right.\!\right)\!,
\end{equation}
where $p(\nlmi{1}|\Nlm{k})$ is the square root of the binomial distribution, $ h_{\Klmk}(\nlmi{1}|\Nlm{k})$ is a $\Klmk$th order discrete Hermite polynomial, and $ a_{\Klmk}(\Nlm{k})$ is a normalization factor (for an expression of these coefficients see App.~\ref{Sec:expr_HO}).  Notice that the one-level Hamiltonians, expressed in the Fock basis, resemble, in the non-interacting limit,  a harmonic oscillator potential  truncated at hard walls. This gives rise to the binomial distribution and Hermite polynomial, as appropriate for such a potential. 
 The corresponding eigenvalues are
\begin{equation}
\label{eq:eigenv_OSC}
  \varepsilon_{k\ell m} = -J_{\ell m}\left(\Nlm{k}-2\Klmk\right) + \ell\hbar\omega\Nlm{k}.
\end{equation}
Because the amplitudes $\cilmk$ resemble the eigenfunctions of the 1D harmonic oscillator potential and the eigenvalues $\varepsilon_{k\ell m}$ are linear in $\Klmk$, the eigenstates (\ref{eq:HO_states}) are said to be harmonic-oscillator-like.

The ground state $|\phi_0\rangle$ is a coherent superposition of atoms in the lowest energy level of the left and right wells. The probability density of the ground state is
\begin{equation}
 \big|c_i^{(0)}\big|^2 = \frac{1}{2^{N}}\left[\frac{N!}{n_{100}^{(i)}!\left(N-n_{100}^{(i)}\right)!}\right]\delta_{N_{00}^{(i)}N},
\end{equation}
with corresponding energy
\begin{equation}
  \varepsilon_0 = -NJ_0.
\end{equation}
This result is readily generalizable to an arbitrary number of energy levels.

\subsection{High barrier: macroscopic superposition states}

We now turn our attention to the opposite, high barrier limit or Fock regime, $\zeta\gtrsim 1 $.   Let us assume first that $J_0=J_1=0$, i.e., the infinite-barrier limit. In this regime Eq.~(\ref{Eq:Uscales}) holds, and it is evident that none of the coefficients $ U^{\ell m}_{\ell' m'} $ can be neglected. Then, the eigenvectors of Hamiltonian~(\ref{eq:two-level}) are not Fock vectors, due to the terms $\left[ \!\left(\!\bhat{j}{1}{0}{\dagger}\!\right)^2\!
  \bhat{j}{1}{1}{}\bhat{j}{1}{-1}{}
  +\hc \right]  $ and $\left[ \!\left(\!\bhat{j}{0}{0}{\dagger}\!\right)^2\!
     \bhat{j}{1}{m''}{}\bhat{j}{1}{-m''}{}
     \!+\!\hc \right]$.  Nevertheless, we can neglect these terms whenever $ 2\triangle E\gg N^2 U_0$. We will justify this criterion in the next section.  Then, the eigenvectors of the resulting Hamiltonian are, indeed, Fock states with eigenvalues:
\begin{align}\label{eq:eigenHB}
\epsilon_k&= \sum_{\ell,m}\Bigg\{E_{\ell} N_{\ell m}^{(k)}+U_{\ell m}^{\ell m}\left[2\left(n_{1\ell m}^{(k)}-\frac{N_{\ell m}^{(k)}}{2}\right)^{2}\right.\nonumber\\
&+\left.N_{\ell m}^{(k)}\left(\frac{N_{\ell m}^{(k)}}{2}-1\right)\right]\Bigg\}+\sum_{j\ell m}\sum_{m'\ne m}2\,n_{j\ell m}^{(k)}n_{j\ell m'}^{(k)}\nonumber\\
&+\sum_{j,m''}4U_{00}^{1m''}n_{j00}^{(k)}n_{j1m'}^{(k)}.
\end{align}

According to Eq.~(\ref{eq:eigenHB}) the number of degenerate eigenstates depends on the occupation of the excited level. For example, for no atoms in the excited level, there are two degenerate eigenstates obeying $ n_{100}^{(k)} = n_{200}^{(k')} $  and $ n_{200}^{(k)} =n_{100}^{(k')}$. For one atom in the excited level with $m=0$ there are four degenerate eigenstates, since the excited atom can be located in any of the two wells, thus giving four combinations.  Let us now consider  $J_0 \ll U_0$ and $J_1 \ll U_0$. Non-degenerate perturbation theory gives, in every case, that the eigenvectors are quasi-degenerate symmetric and antisymmetric combinations of the corresponding Fock vectors (see App.~\ref{Sec:HighBarrier}).
For the particular cases in which all atoms occupy the same level, and for which the angular momentum of each atom is oriented along the $z$-axis, that is, $m = \pm\ell$, the eigenstates are
\begin{equation}\label{eq:MS}
  \phiMS{\ell}{\pm\ell}{\pm}{\nu} = \MS{\ell}{\pm\ell}{\nu},
\end{equation}
for $0\leq \nu < N/2$; we have neglected terms on the order of $(J_0/U_0)^{N-2\nu}$ and smaller.  Here
\begin{align}
  \MS{\ell}{m}{\nu}
  &\equiv\frac{e^{i\varphi_0}}{\sqrt{2}}\left[\frac{1}{\sqrt{\nu!(N-\nu)!}}
    \left(\bhat{1}{\ell}{m}{\dagger}\right)^{\nu}
    \left(\bhat{2}{\ell}{m}{\dagger}\right)^{N-\nu}\right.\nonumber\\
	 &\pm
   \left.\frac{1}{\sqrt{\nu!(N-\nu)!}}
   \left(\bhat{1}{\ell}{m}{\dagger}\right)^{N-\nu}
   \left(\bhat{2}{\ell}{m}{\dagger}\right)^{\nu}
  \right]|0\rangle,
\end{align}
is an MS state in which $\nu$ and $N-\nu$ atoms simultaneously occupy the $m$th orbital state of the $\ell$th energy level of both wells.  Here $ \varphi_0$ is the usual arbitrary phase associated with vectors in a Hilbert space. We will set $ \varphi_0=0$ for the rest of this Article.  The special case $\nu=0$ represents an \emph{extreme} MS state in which all $N$ atoms simultaneously occupy the left and right wells.  These MS states can be either symmetric $(+)$ or antisymmetric $(-)$.

The eigenstates (\ref{eq:MS}) occur in nearly degenerate pairs of symmetric and antisymmetric MS states.  The level splitting between the states $\phiMS{\ell}{\pm\ell}{-}{\nu}$ and $\phiMS{\ell}{\pm\ell}{+}{\nu}$ is $\Delta\varepsilon_{\ell}(\nu)$ where
\begin{equation}
  \Delta\varepsilon_{\ell}(\nu) =
  \frac{4U_{\ell m}^{\ell m}[J_{\ell m}/(2U_{\ell m}^{\ell m})]^{N-2\nu}(N-\nu)!}
  {\nu![(N-2\nu-1)!]^2},
\end{equation}
up to $(N-2\nu)$th order in $J_0/U_0$. In agreement with the rotational symmetries of the potential, the states $\phiMS{1}{+1}{\pm}{\nu}$ and $\phiMS{1}{-1}{\pm}{\nu}$ are degenerate.  On the other hand, the energy difference between states $\phiMS{0}{0}{\pm}{\nu}$ and $\phiMS{1}{\pm1}{\pm}{\nu}$ is on the order of $N\triangle E$ when $N U^0\ll 2\triangle E$.

Since  $J_1>J_0$, it is also possible that $J_0 \ll U_0$ but $J_1 > U_0$. Then, atoms in the bottom level behave as in the Josephson regime while the ones in the excited level behave as in the Fock regime. We will show numerical examples of these mixed regime in Sec.~\ref{Sec:Numerics}.

\subsection{Shadows of macroscopic superposition states}

Let us turn now our attention to the effects of the coupling between energy levels  in the high barrier limit or Fock regime, $J_{0}\ll U_{0}$ and  $J_1\ll U_0$.   Let us consider the hopping terms and the interaction terms that account for same-site interlevel hopping in Hamiltonian~(\ref{eq:two-level}) as a perturbation to the decoupled Hamiltonian. As shown in App.~\ref{Sec:Interlevel_pert} for the $N+1$ eigenvectors with zero occupation of the excited level, the first order approximation to the eigenvector shows coupling to states with different number of atoms in the excited band.  
These couplings are associated with the destruction of two atoms in the lower level and creation of two atoms in the first level with $m=0$ or  one with $m=1$ and the other with $m=-1$. The corresponding coefficients are negligible as far as $2\triangle E\gg NU_{0}$. Nevertheless, if $2\triangle E$ is comparable to $NU_{0}$, Fock vectors with non-zero occupation of the excited levels are coupled to the MS states. We call these coupled vectors  shadows of the MS states $|\phi_{\pm}^{(0)};n_L\rangle$. Similar results hold for MS states with nonzero occupation of the excited level. Coupling between different levels in asymmetric double wells or optical lattices plays a fundamental role in far-from-equilibrium dynamics showing Landau-Zener (LZ) coupling~\cite{Smith:2009,Chen:2010}. Here we focus on statics and on the symmetric case, leaving for future work the study of how LZ coupling between different wells in asymmetric double well potentials is modified in this regime. 

\section{Bounds on the Use of a One- and Two-Level Approximation}
\label{Sec:bounds}
We have distinguished two main regimes, the Josephson regime, in which the eigenstates are HO-like and, the Fock regime, in which MS states can be found. The  Josephson regime is characterized by: 
\begin{subequations}
 \begin{equation}
 \eta_{\mathrm{Jos},\,\ell}\equiv\frac{NU_{\ell m}^{\ell m}}{J_{\ell}}\ll 1.
\label{eq:CA}
\end{equation}
The Fock regime is characterized by:
\begin{equation}
 \eta_{\mathrm{Fock},\,\ell}\equiv\frac{J_{\ell}}{U_{\ell m}^{\ell m}}\ll 1.
\label{eq:CB}
\end{equation}
\end{subequations}

Notice that, as stated above, these criteria should be evaluated for both levels. Then, it is possible that the Fock regime holds for atoms in the bottom level, while the Josephson regime holds for atoms in the excited level. Since $J_1>J_0$ the contrary is not true. Hence, in general, we distinguish three regimes: the Josephson regime, the Fock regime, and the mixed regime. In the first two regimes,  the corresponding criterion holds for atoms in both levels. In the latter,  the Fock  criterion holds for atoms in the bottom level, while the Josephson criterion is satisfied for atoms in the excited level.  Let us show, for these regimes, the bounds on the one- and two- level approximations. With our choice of indexing states, the one-level approximation corresponds to truncating the size of the Hilbert space to $N+1$.  Then, the bounds we present below will describe the regime in which this truncation is valid.

Let us consider first the Josephson regime. The energy levels are coupled by the interaction energy
$U^{\ell m}_{\ell'm'} $, $\ell\ne\ell' $. In this regime all coefficients $ U^{\ell m}_{\ell'm'}$ are small, and then,  the coupling between levels is weak.  Energy levels only become completely decoupled when $U^{\ell m}_{\ell'm'}=0 $. However, let us show that eigenvalue crossings are induced by the presence of the excited level.
Let us assume that $U_{0}=0$, which implies that $ U^{\ell m}_{\ell'm'}=0$. According to Eq.~(\ref{eq:eigenv_OSC}),
the maximum of the eigenvalues for the first $N+1$ eigenstates with no occupation of the excited level coincides with the minimum of the eigenvalues of the states with one atom in the excited level, if
\begin{equation}\label{eq:CC}
\chi_{\mathrm{Jos}}\equiv\frac{\triangle E}{ J_{0}\,(2N-1)+J_1}=1,
\end{equation}
this being the criterion that determines the first eigenvalue crossing in this regime. For $ \chi_{\mathrm{Jos}}>1$ no crossing occurs. Moreover, for
\begin{equation}\label{eq:CD}
\chi_{\mathrm{Jos, gs}}=\frac{\triangle E}{ J_{1}-J_0}=1,
\end{equation}
the first crossing involving the ground state occurs, i.e., the ground state shows non-zero occupation of the excited level if $\chi_{\mathrm{Jos, gs}}<1$.

Analogously, in the Fock regime,  the first eigenvalue crossing occurs when the condition
\begin{equation}\label{eq:CE}
\chi_{\mathrm{Fock}}\equiv\frac{U_{00}^{00}(N^2+2N-3)}{2\left(\triangle E-2U_{00}^{10}(N-1)\right)}=1
\end{equation}
is met. This condition is obtained equating the maximum eigenvalue given by Eq.~(\ref{eq:eigenHB}) for states with no occupation of the excited level, to the minimum eigenvalue given by this equations for states  with one atom in the excited level. For $\chi_{\mathrm{Fock}}<1$ no crossing takes place. For large $N$ this criterion can be approximated by $ N^2U_0\sim2\triangle E$. On the other hand,  the first crossing involving the ground state occurs when the condition:
\begin{equation}\label{eq:CF}
\chi_{\mathrm{Fock, gs}}=\frac{(3/2)U_{00}^{00}(N-1)-2U_{00}^{10}(N-1)}{\triangle E}=1
\end{equation}
is satisfied. For $\chi_{\mathrm{Fock, gs}}<1$  the ground state shows non-zero occupation of the excited level. For large $N$ this criterion is $ NU_0\sim2\triangle E$.

If we consider Eq.~(\ref{Eq:Uscales}), in the high barrier limit, the criteria (\ref{eq:CE}) and (\ref{eq:CF}) turns into:
\begin{equation}\label{eq:CG}
\chi_{\mathrm{Fock}}^{\mathrm{approx}}=\frac{U_0(N^2-1)}{2\triangle E}=1
\end{equation}
and
\begin{equation}\label{eq:CH}
\chi_{\mathrm{Fock, gs}}^{\mathrm{approx}}=\frac{U_0(N-1)}{2\triangle E}=1.
\end{equation}

For completeness, let us consider the following criterion:
\begin{eqnarray}\label{eq:CI}
\chi_{\mathrm{shadow}}&\equiv& \frac{\sqrt{2}U_{00}^{11}\sqrt{N\left(N-1\right)} }{U_0(4N-6)-2 U_{10}^{10}-8U_{00}^{10}(N-2)-2\triangle E}.\nonumber\\
\end{eqnarray}
In the Fock regime, the expression of the MS states in terms of the Fock basis gives  only two relevant coefficients, those corresponding to the two Fock vectors that are superimposed, as detailed in Sec.~\ref{Sec:CharEig}. But, as the interactions are increased the coefficients corresponding to the shadows of MS states are more relevant. The greatest of these coefficients corresponds to a Fock vector with two atoms in the excited level with $m=0$. The criterion~(\ref{eq:CI}) is the expression of this coefficient, which is obtained in App.~\ref{Sec:Interlevel_pert} using perturbation theory.  Hence, for small  $\chi_{\mathrm{shadow}}$ we can neglect the interlevel coupling, since all coupling to other Fock vectors will be negligible. If we consider the relations given by Eq.~(\ref{Eq:Uscales}), this criterion can be written as:
\begin{equation}\label{eq:CJ}
\chi_{\mathrm{shadow}}^{\mathrm{approx}}=\frac{U_0\sqrt{N\left(N-1\right)}}{\sqrt{2}\triangle E\left(U_0/2\triangle E-2\right)}.
\end{equation}

Moreover, the single particle eight-mode basis turns to not be  appropriate to express the field operators if the barrier height is smaller than the energy gap between levels $\triangle E$, i.e., if 
\begin{equation}\label{eq:CM}
\chi_{\triangle E}\equiv\frac{V_0}{\triangle E}<1.
\end{equation}
Finally, the criterion 
\begin{equation}\label{eq:CK}
\chi_{\mathrm{weak}}=\frac{N^{\frac{1}{3}}U_0}{2\triangle E}\ll 1
\end{equation}
has to be met to account for the weakly interacting gas condition, Eq.~(\ref{eq:diluteness}). Equation~(\ref{eq:CK}) is obtained using the analytical form of $U_0$ and $\triangle E$ to elliminate $\aho/\as$ in Eq.~(\ref{eq:diluteness}), taking into account all the scalings performed.  

Therefore, the ground state shows occupation of the excited level when  criteria~(\ref{eq:CD}) and (\ref{eq:CF}),  or its high barrier version (\ref{eq:CH}), are fulfilled. 
 Moreover, we are interested in describing cat-like MS states which are typically excited eigenstates. Therefore, a characterization of eigenvalue crossings of energies other than the ground state is also relevant. These crossings appear when criteria~(\ref{eq:CC}) and (\ref{eq:CE}), or  (\ref{eq:CG}), are satisfied. Also, criterion~(\ref{eq:CI}),  or  (\ref{eq:CJ}), indicates the presence of shadows of MS states.  It is important to notice that, for large $N$, both criteria~(\ref{eq:CF}) and~(\ref{eq:CI}) becomes $NU/2\triangle E$, while criterion (\ref{eq:CE}) becomes $N^2U/2\triangle E$.  Then, concerning MS states, the excited level plays a relevant role insofar $N^2U  \sim 2\triangle E$. Furthermore,  when $NU \sim 2\triangle E$  the ground state shows occupation of excited levels and the coupling between  levels is non-negligible. Then, for large $N$, the use of single-particle wavefunctions and only a few energy levels is appropriate to the regime
\begin{equation}\label{eq:CL}
  \chi_{\mathrm{model}}=\frac{N U_0}{2\triangle E}\ll 1.
\end{equation}
If this condition is not met, our approximation is inaccurate and alternative treatments become necessary~\cite{Masiello:2005,Streltsov:2006}. Notice that this criterion is identical to the regime for which mean field theory is valid. 
 Finally, criteria~(\ref{eq:CM}) and (\ref{eq:CK}) are also two limiting criteria for the model.

\section{Numerical Results}
\label{Sec:Numerics}

\begin{figure}[!t]\center
\vspace*{-.5cm}
\includegraphics[width=\columnwidth]{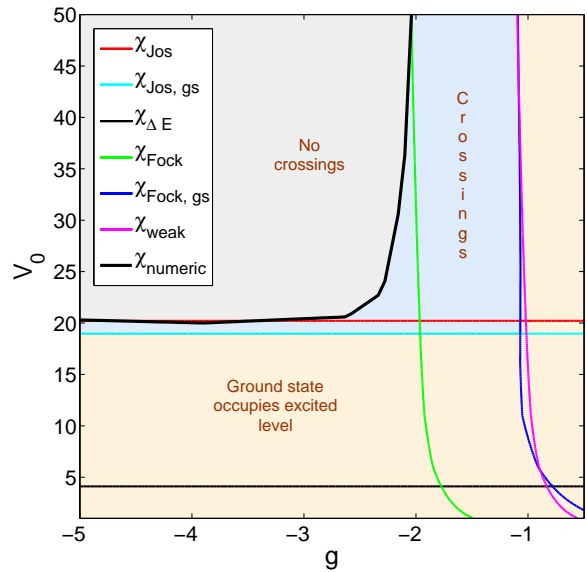}
\vspace*{-1.35cm}
  \caption{(Color online) {\it  Characterization of the crossings and validity of the model in the $V_0$-$g$ plane (both dimensionless), for the Duffing potential.} Different criteria that characterize the first crossing and the first crossing involving the ground state, as well as the limits of validity of the model for $N=8$ atoms. Three main areas are distinguished: (i) the area in which no crossing occurs; (ii) the area in which eigenstates with occupation of the excited level emerge among the first $N+1$ eigenvectors; (iii) and the area in which even the ground state shows occupation of the excited level. The one-level approximation is valid in the area (i). \label{fig:Vgplane}}
\end{figure}

Let us use the previous criteria to completely demarcate the different regimes in the   $V_0$-$g$ plane. At every point of this plane different values of the interaction coefficients,  $ U^{\ell m}_{\ell' m'}$, the hopping coefficients, $ J_{\ell m}$, as well as the energy level spacing, $\triangle E$, are obtained. Since all criteria depend on these parameters, we can draw in this plane the curves for which the different criteria are satisfied, thus determining different regions in which the eigenstates have been characterized.  We will illustrate the results with examples obtained after exact diagonalization of the Hamiltonian~(\ref{eq:two-level}). Also, we will determine in that plane the limits of validity of our model.

\begin{figure}[!t]\center
\vspace*{-.75cm}
\includegraphics[width=\columnwidth]{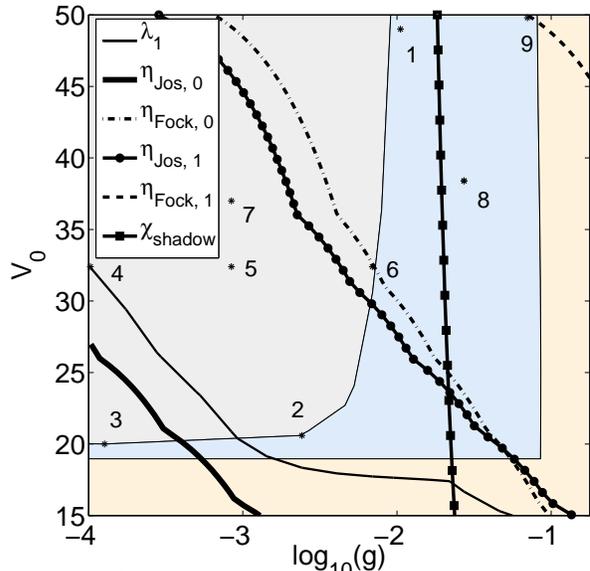} 
\vspace*{-1.cm}
  \caption{(Color online)   {\it  Characterization of the different regimes in the $V_0$-$g$ plane (both dimensionless).} Different criteria that characterize the Josephson and Fock regimes for both levels when $N=8$. The numbered points represent the examples given in subsequent figures.  For completeness, the three regions distinguished in Fig.~\ref{fig:Vgplane} are also represented.   \label{fig:Vgplaneb}}
\end{figure}

\begin{figure}
\begin{tabular}{c}
\vspace*{-0.75cm}\tabularnewline
 \includegraphics[width=7.25cm]{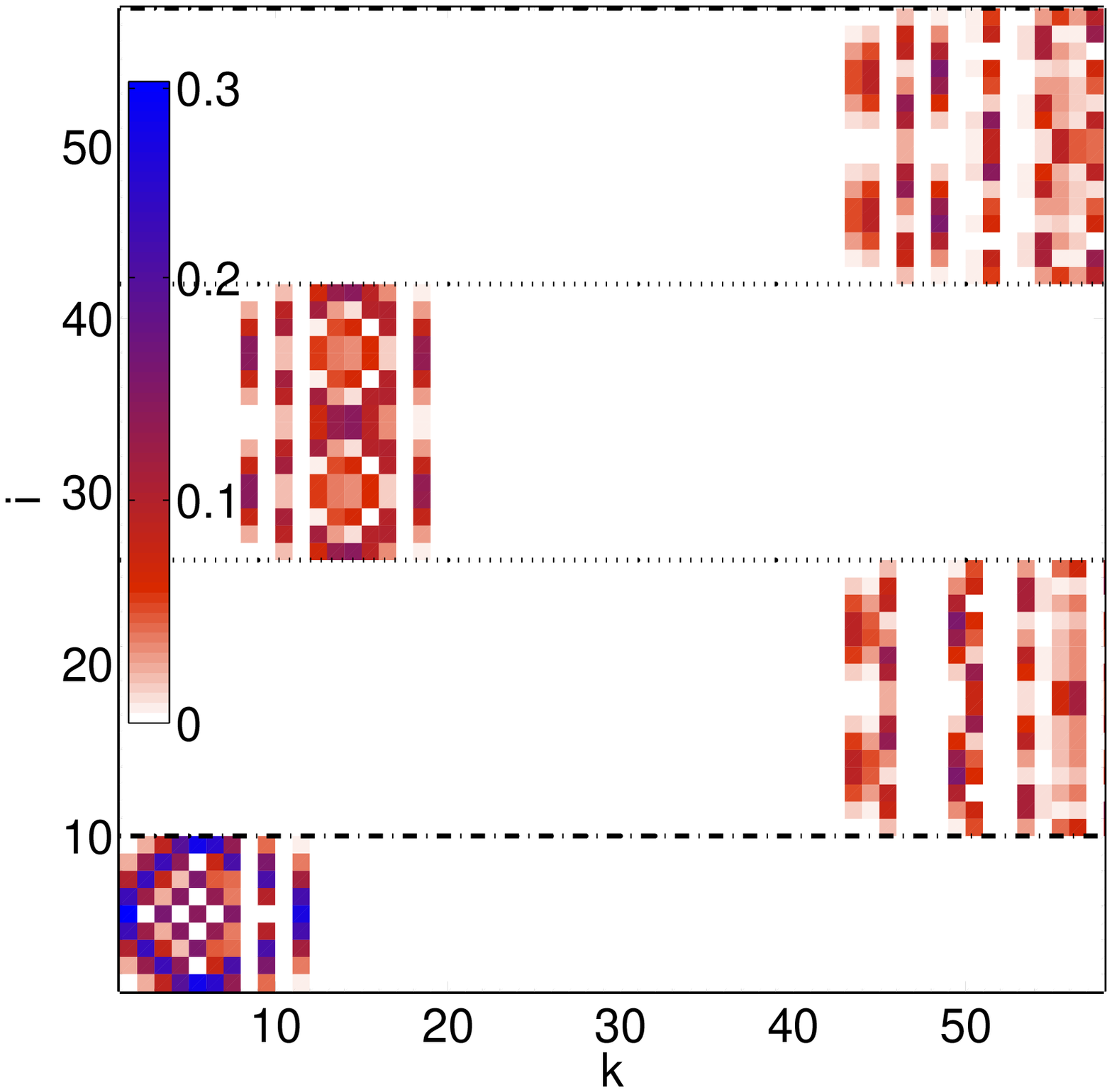} \tabularnewline
\vspace*{-6.5cm}\tabularnewline
(a) \tabularnewline
\vspace*{4.75cm}\tabularnewline
  \includegraphics[width=7.25cm]{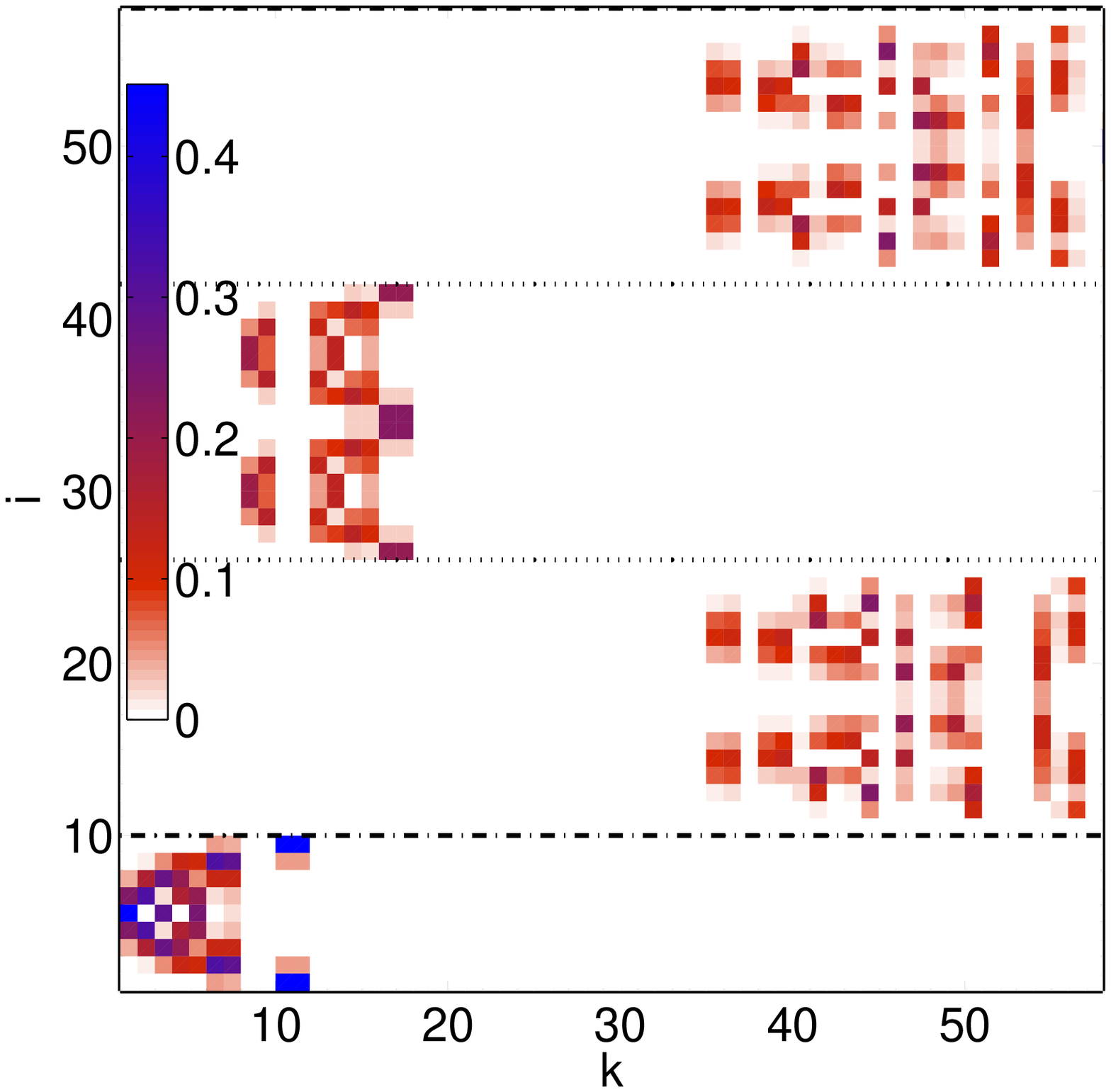}\tabularnewline
\vspace*{-6.35cm}\tabularnewline
  (b)\tabularnewline
\vspace*{4.75cm}\tabularnewline
  \includegraphics[width=7.25cm]{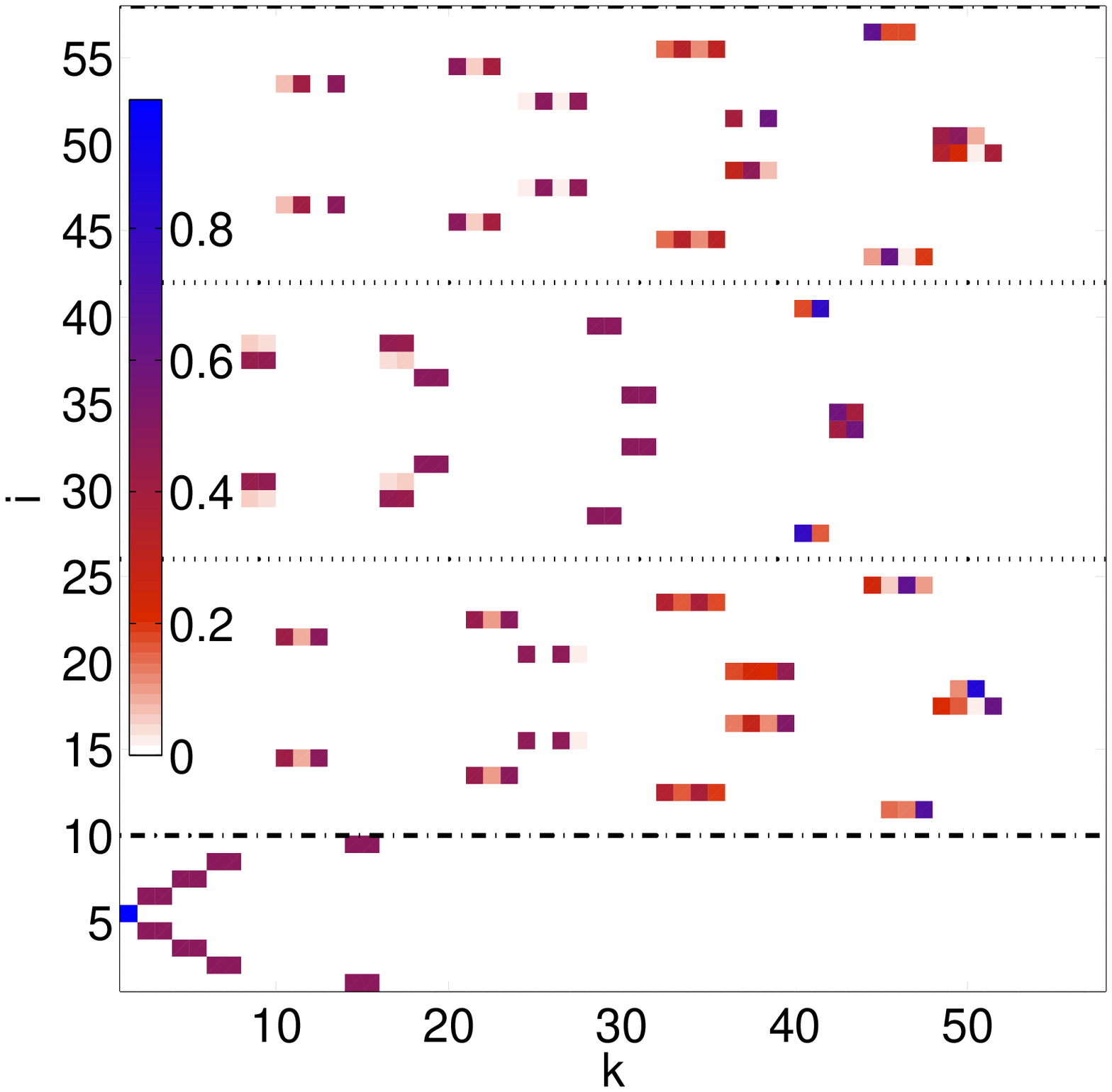}
\vspace*{-6.05cm}\tabularnewline
  (c)\tabularnewline
\vspace*{5cm}\tabularnewline
\end{tabular}
  \caption{(Color online)  {\it Eigenstates for  three examples of the first crossing in the Josephson, intermediate, and Fock regime.} Probability amplitudes $|\cilmk|^2$ of  the first  $N+1+6N$ eigenstates for (a) $V_0=20$; $g\approx10^{-4}$, (b) $V_0=20.6$; $g\approx2\times10^{-3}$, and (c) $V_0=49$; $g\approx 10^{-2}$, when $N=8$ atoms.  Above the dashed line one atom
occupies the excited level. Dotted lines separate the Fock vectors for which this atom shows $m=-1,0,1$, respectively.   \label{fig:ho1c}}
\end{figure}

In figure~\ref{fig:Vgplane} we represent, for $N=8$  atoms, the curves in the $V_0$-$g$  plane for which the criteria that characterize the first crossing,  Eq.~(\ref{eq:CC}) and  Eq.~(\ref{eq:CE}), are fulfilled. These criteria are valid for large $g$ and $V_0$, respectively. On the other hand, an intermediate regime arises between the Josephson and Fock regimes for which the interaction coefficient is comparable to the hopping coefficient. Since the results obtained theoretically are not valid in this intermediate regime we have performed exact diagonalization of the Hamiltonian~(\ref{eq:two-level}) to determine the pairs $V_0$-$g$ that lead to the first crossing. The corresponding interpolated curve is presented as the criterion $\chi_{\mathrm{numeric}}$ in the figure. In Fig.~\ref{fig:Vgplaneb}, we also represent  this curve.  Points 1, 2, and 3 in this figure correspond to Figs.~\ref{fig:ho1c} (a), (b), and (c), where the coefficients $|\cilmk|^2$ for the first $N+1+6N$ eigenvectors are represented, thus showing the first crossing in the first $N+1$ eigenvectors.
Hence, there is no crossing in the region to the left of this curve. We also represent in Fig.~\ref{fig:Vgplane}  the curve for which criteria  for the first crossing involving the ground state in the Josephson    [Eq.~(\ref{eq:CD})] and Fock [Eq.~(\ref{eq:CF})] regimes are satisfied. In the region between these curves and the one given by  $\chi_{\mathrm{numeric}}$ the crossings do not involve the ground state. Finally, the curves associated with criterion~(\ref{eq:CK}) and  criterion~(\ref{eq:CM})  demarcate the limits of validity of the model.  

Once these three main regions have been defined, let us show in this plane the different regimes in which the individual eigenstates have been characterized. We represent in Fig.~\ref{fig:Vgplaneb}, for the bottom level and for the excited level with $m=0$,  the criterion for the limit of the Josephson regime, Eq.~(\ref{eq:CA}),  for $\eta_{\mathrm{Jos},\,\ell}=0.1$, $\ell=0,1$. Similarly, we represent the criterion for the limit of the Fock regime, Eq.~(\ref{eq:CB}),  for $\eta_{\mathrm{Fock},\,\ell}=0.1$, $\ell=0,1$.  Cusps in both curves are an artifact associated with the resolution of the interpolation between the points in which we have numerically calculated the eigenfunctions and all the coefficients. 

\begin{figure}
\begin{tabular}{c}
\vspace*{-1.15cm}\tabularnewline
 \includegraphics[width=7.15cm]{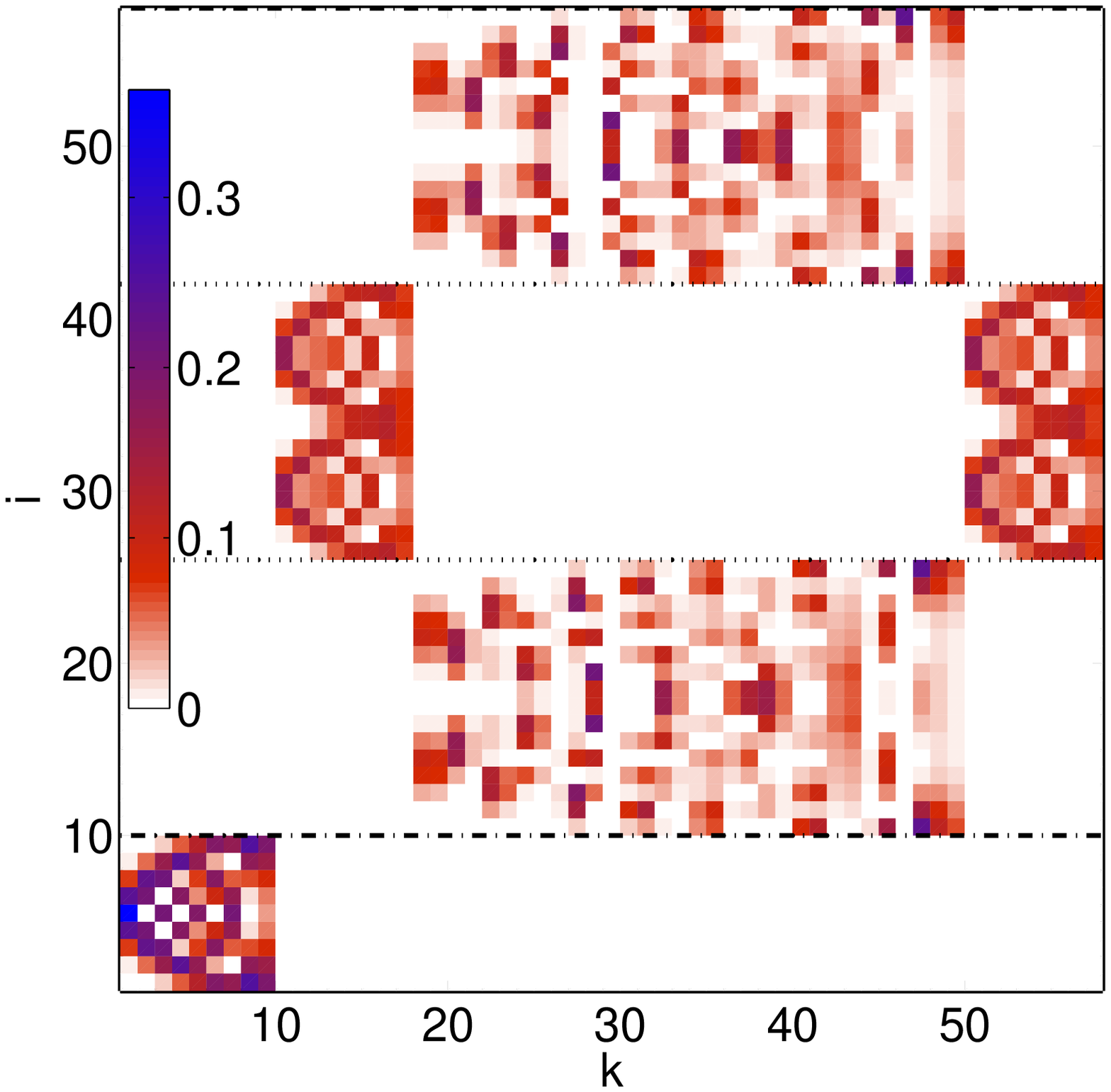} \tabularnewline
\vspace*{-6.25cm}\tabularnewline
\hspace*{-3.5cm}(a) \tabularnewline
\vspace*{5.25cm}\tabularnewline
  \includegraphics[width=7.15cm]{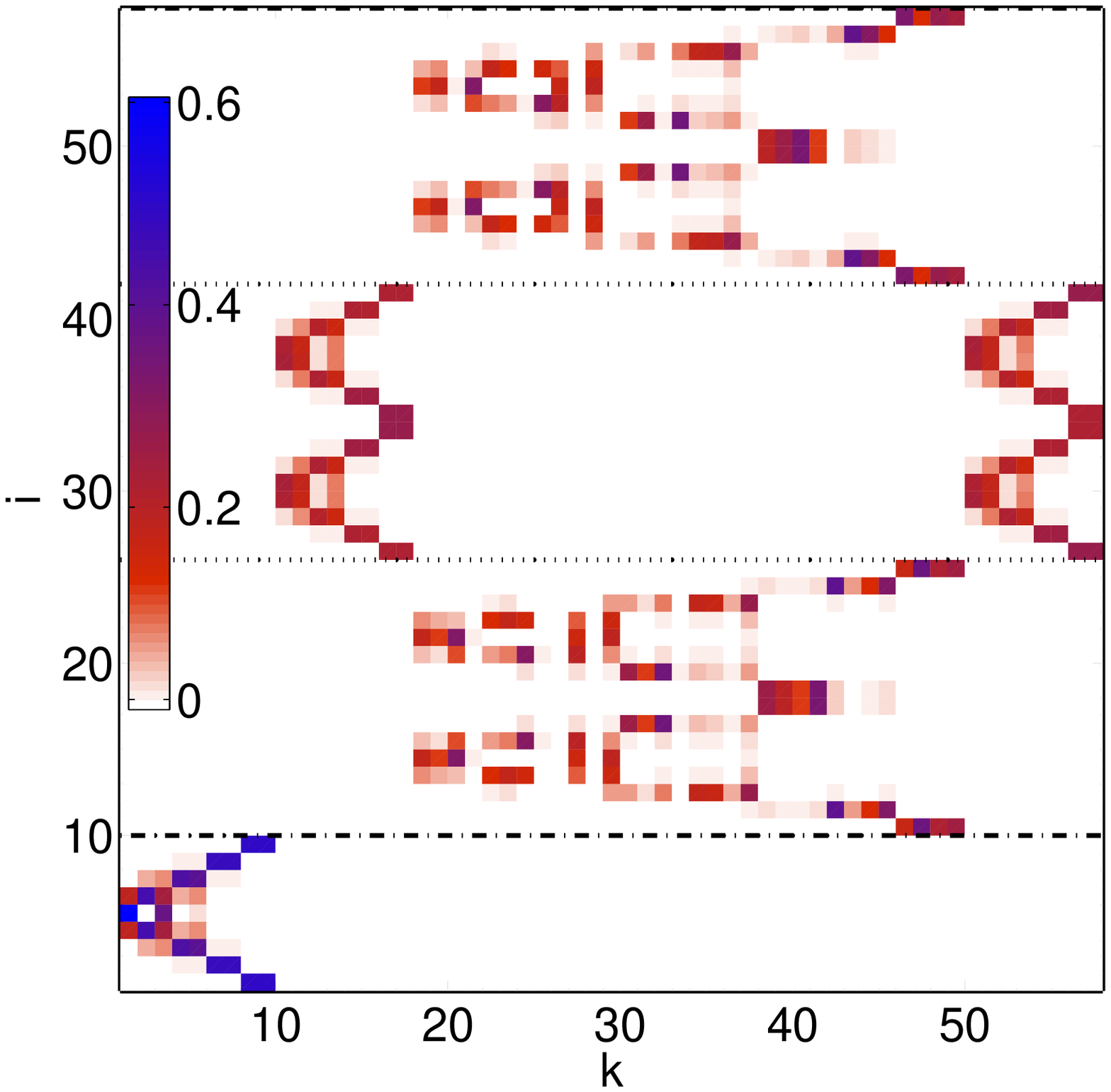}\tabularnewline
\vspace*{-6.15cm}\tabularnewline
\hspace*{-3.5cm}  (b)\tabularnewline
\vspace*{5.15cm}\tabularnewline
  \includegraphics[width=7.15cm]{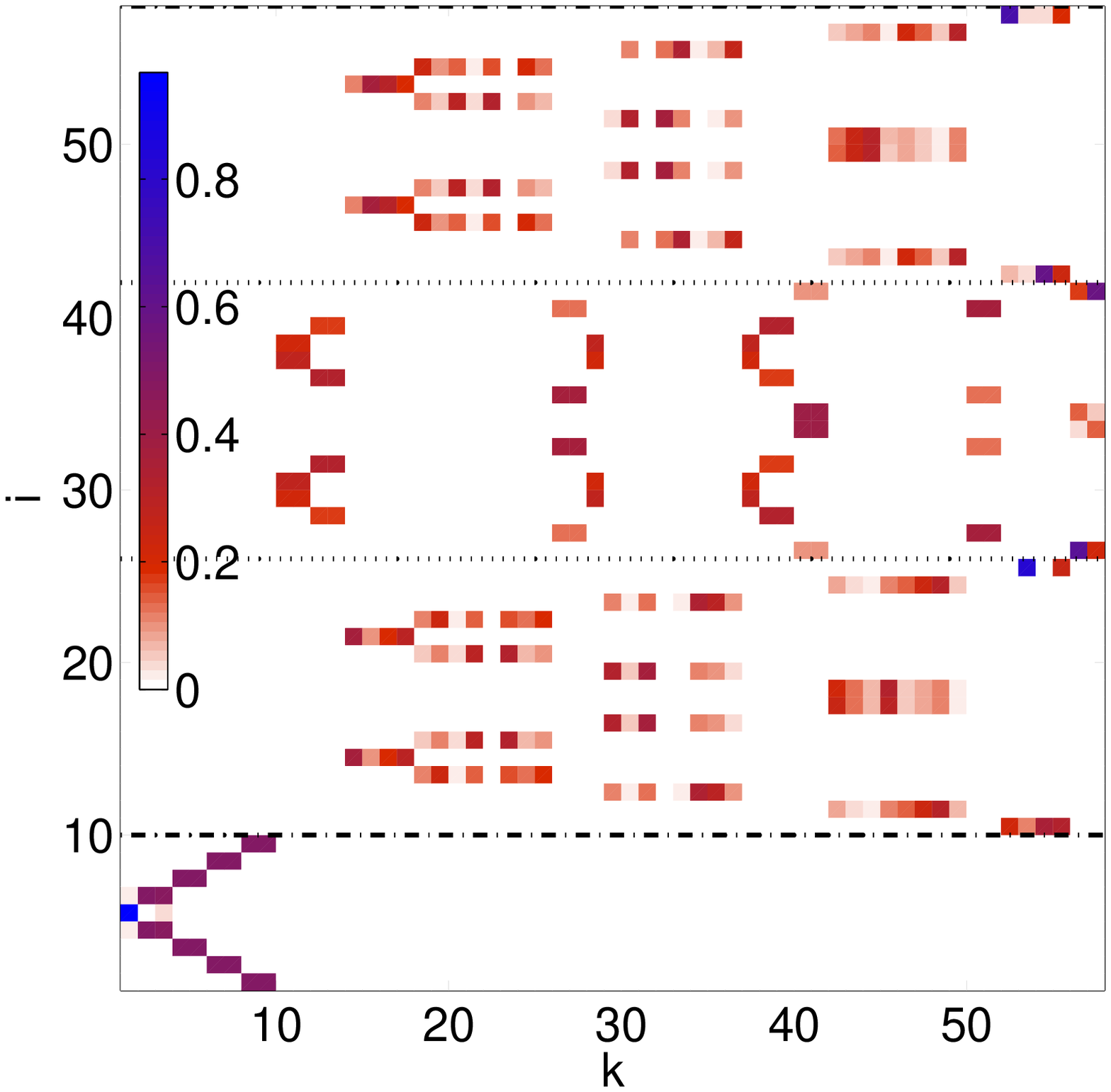}
\vspace*{-5.8cm}\tabularnewline
 \hspace*{-3.5cm} (c)\tabularnewline
\vspace*{5.cm}\tabularnewline
\end{tabular}
  \caption{(Color online)  {\it Eigenstates for  three typical examples of  the eigenvectors from the Josephson, intermediate, and Fock regime.}  Probability amplitudes $|\cilmk|^2$ for  the first $N+1+6N$  eigenstates for $N=8$ atoms, $V_0=32.4$, and  (a)   $g\approx 10^{-4}$, (b)  $g\approx 8\times10^{-4}$, and (c) $g\approx 7\times10^{-3}$. Above the dashed line one atom
occupies the excited level. Dotted lines separate the Fock vectors for which this atom shows $m=-1,0,1$, respectively.  \label{fig:ho1}}
\end{figure}

To further  characterize these regimes, we use the eigenvalues of the single particle density matrix, whose elements are defined as:
$$
\rho_{ij}=\langle \Psi_0 |\hat{b}_{i}^{\dagger}\hat{b}_j | \Psi_0 \rangle,
$$
where $\Psi_0$ is the ground state. Notice that, since the indices $j \ell m$ permit one to run over all combinations $j$, $\ell$, and $m$, this matrix is of dimension eight. The curve $\lambda_1$ represents the pairs $V_0$-$g$ for which the largest eigenvalue takes the value $\lambda_1=  0.85$. Also, the second eigenvalue starts to grow to the right of this curve. Thus, this curve defines the limit of the Josephson regime for the bottom level. In Fig.~\ref{fig:ho1} (a), (b), and (c) we represent examples of the coefficients $|\cilmk|^2$ for the first $N+1+6N$ Fock vectors and eigenstates in the Josephson, intermediate, and Fock regimes, respectively. These examples are represented in Fig.~\ref{fig:Vgplaneb} as points 4, 5, and 6, respectively.

\begin{figure}
\begin{tabular}{c}
\vspace*{-1.15cm}\tabularnewline
  \includegraphics[width=7.5cm]{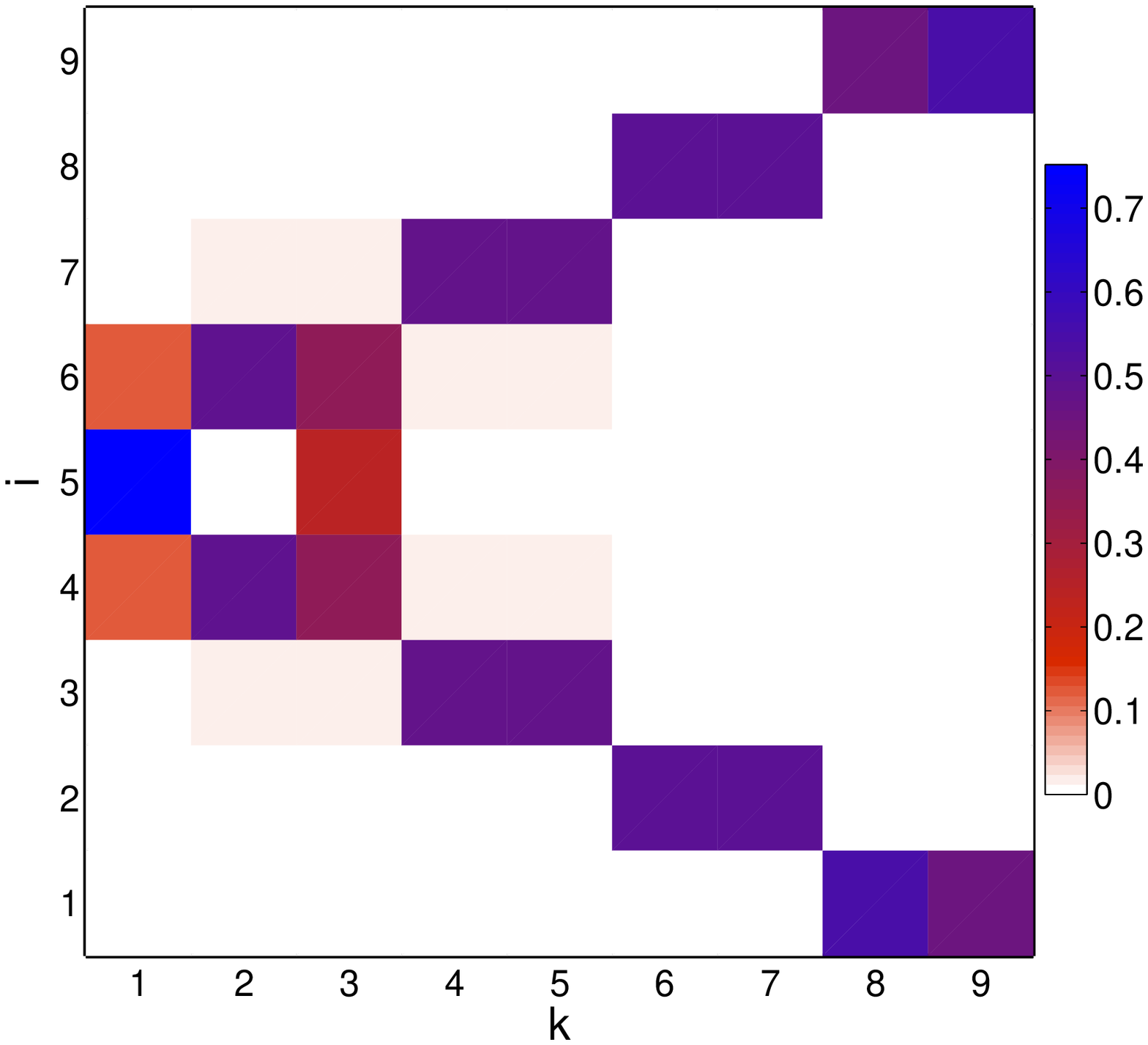}
\tabularnewline
\vspace*{-6.5cm}\tabularnewline
\hspace*{-4.75cm}(a) \tabularnewline
\vspace*{5.5cm}\tabularnewline
\includegraphics[width=7.5cm]{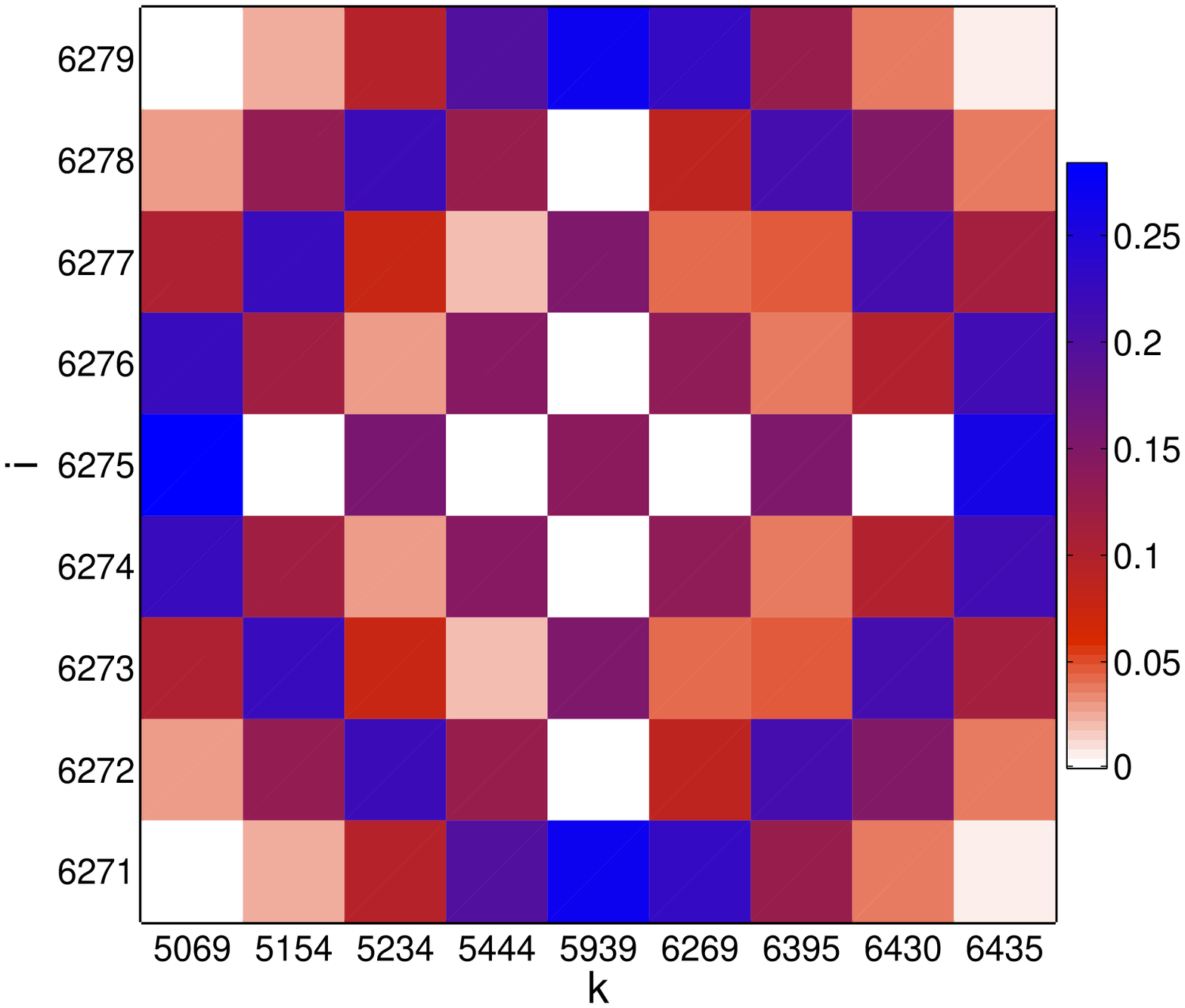}
\vspace*{-6.1cm}\tabularnewline
\hspace*{-4.75cm}(b) \tabularnewline
\vspace*{5.15cm}\tabularnewline
\end{tabular}
  \caption{(Color online)  {\it Eigenstates for the mixed regime.} Probability amplitudes $|\cilmk|^2$ (a)  for  the first $N+1$ eigenstates, in which only  the bottom level is occupied and (b)  for all eigenstates corresponding to occupation only of the excited level with $m=0$ when $N=8$ atoms, $V_0=37$, and $g\approx 10^{-3}$.  Notice that,  while the former eigenstates are MS states, the latter are HO-like states. \label{fig:mix}}
\end{figure}

In Fig.~\ref{fig:mix} we show that the example corresponding to point 7 in Fig.~\ref{fig:Vgplaneb}  belongs to the mixed regime. Hence, the first $N+1$ eigenstates, which show occupation of only  the bottom level, are MS states. Conversely, all eigenstates corresponding to occupation of only  the excited level with $m=0$ behave as HO states. Notice that the latter are not consecutive eigenstates, since there are many crossings in the excited level that we have not considered in detail. 

\begin{figure*}
\begin{tabular}{cc}
\vspace*{-.25cm} & \tabularnewline
  \includegraphics[width=8.5cm]{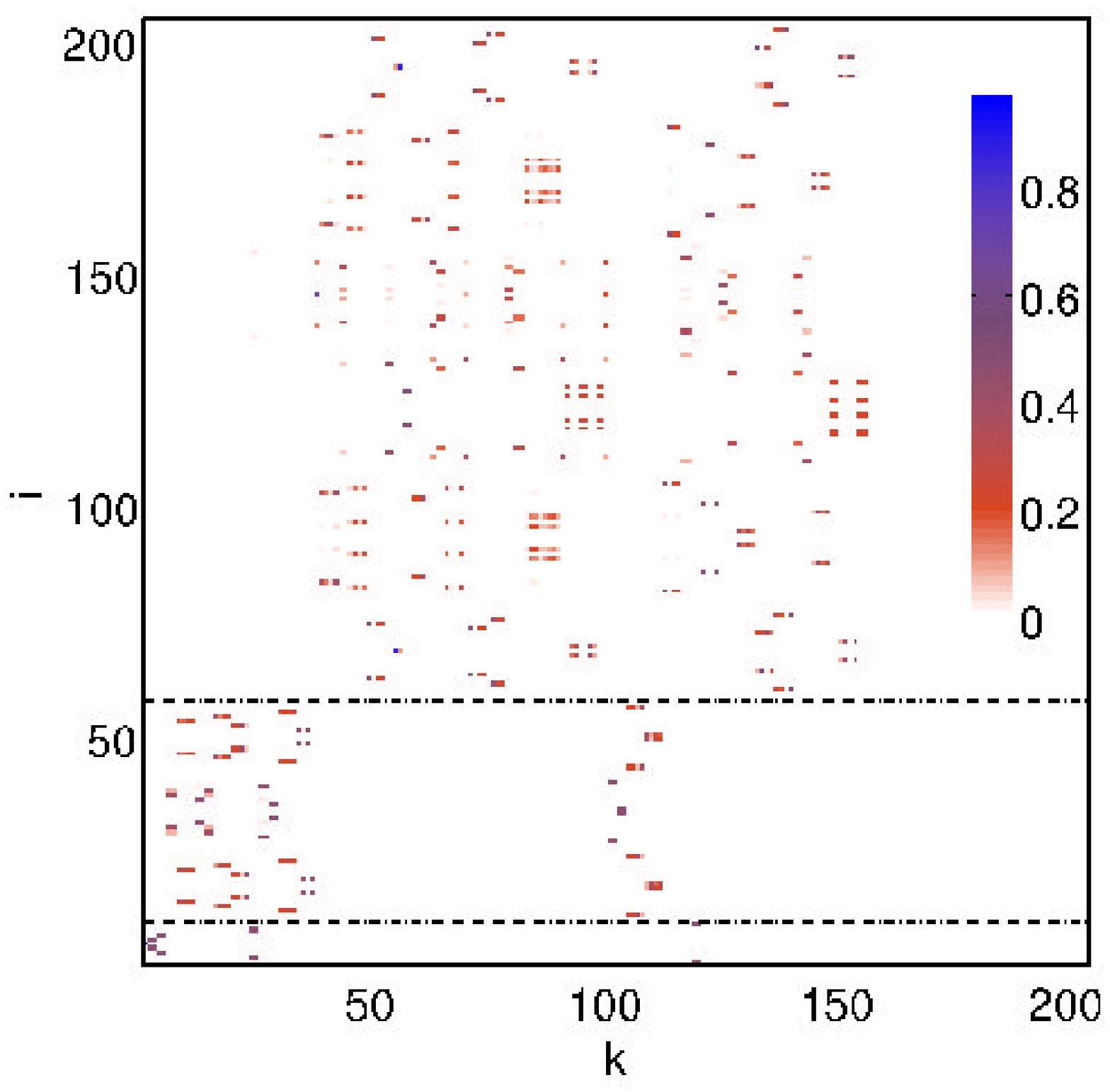}
& \includegraphics[width=8.5cm]{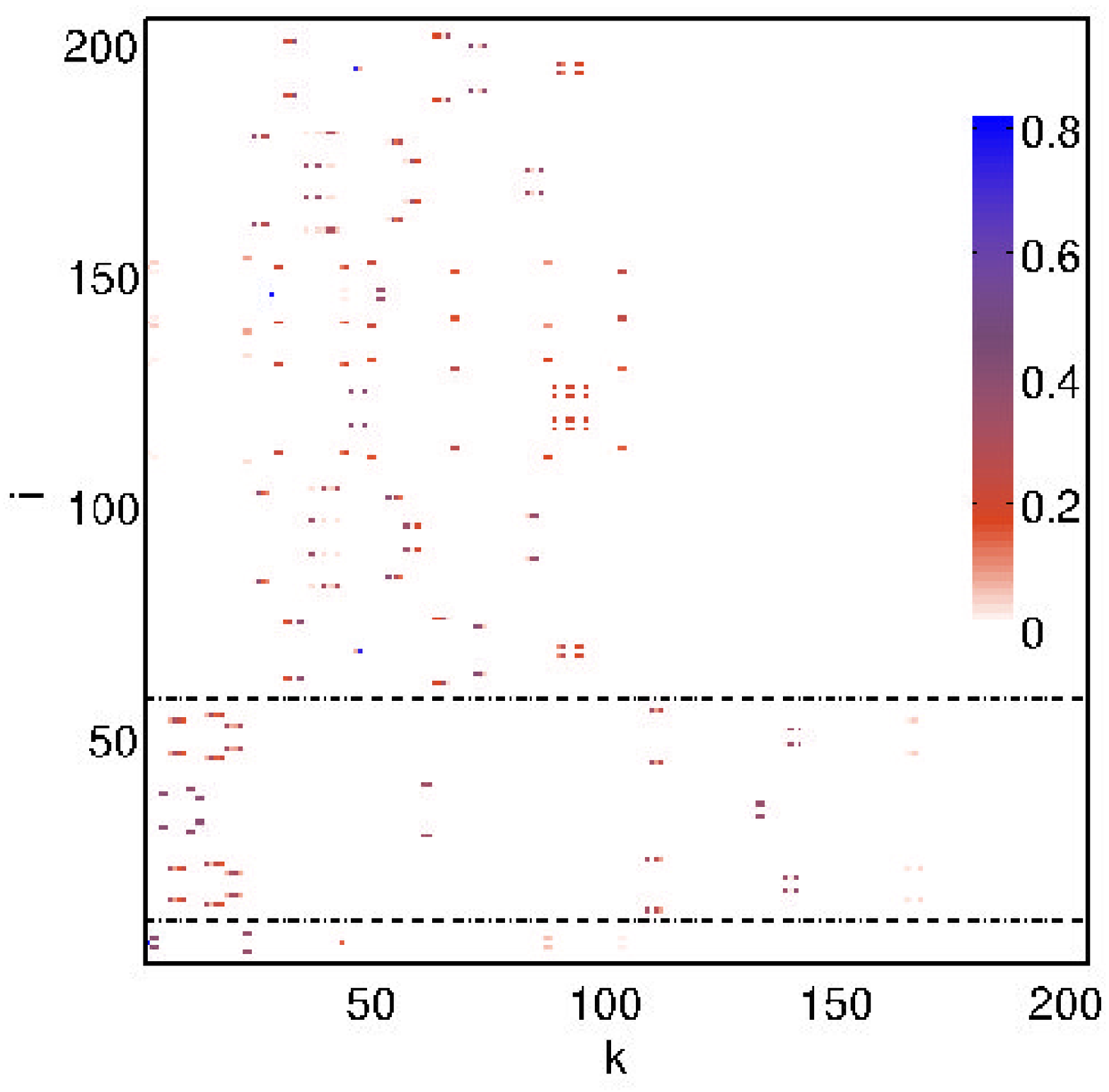} \tabularnewline
\vspace*{-8.5cm} & \tabularnewline
(a) & (c)\tabularnewline
\vspace*{7.9cm} & \tabularnewline
(b) & (d)\tabularnewline
\vspace*{-2.5cm} & \tabularnewline
\hspace*{-1cm} \includegraphics[width=8cm]{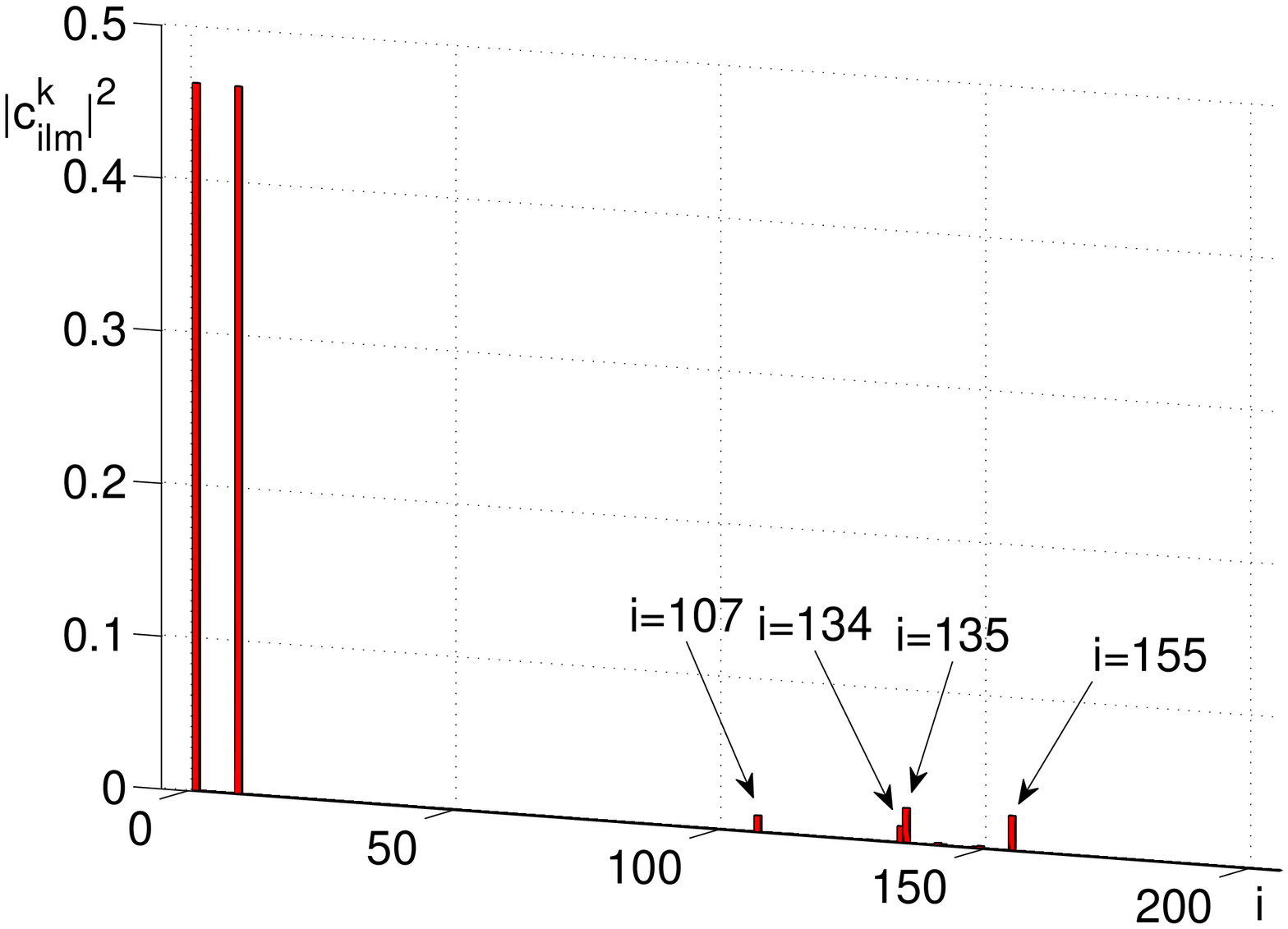}
& \includegraphics[width=8cm]{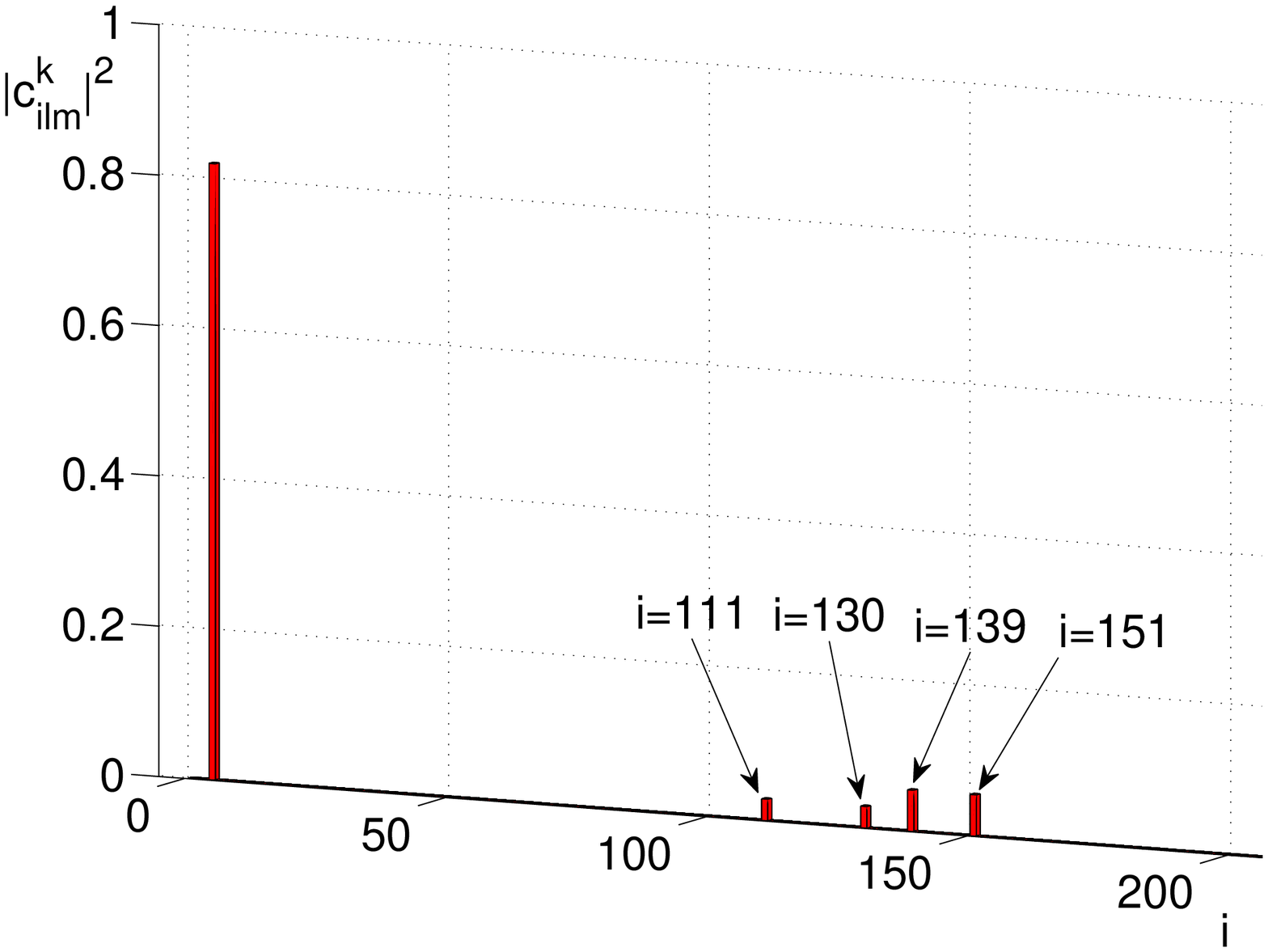}\tabularnewline 
\vspace*{-2.5cm} & \tabularnewline
\end{tabular}
\caption{(Color online)  {\it Shadows of excited MS states and of the ground state.}   (a) Probability amplitudes $|\cilmk|^2$ for the first $(N+1)+6N+21(N-1)$ Fock vectors and eigenstates when $V_0=38.4$, $g\approx 3\times10^{-2}$,   and $N=8$ atoms. One atom
occupies the excited level above the dashed line  at $i = N+1$. Two atoms
occupy the excited level above the second dashed line at $i = N+1+6N$. (b) Coefficients for the 120-th excited state, the most excited state with all atoms in the bottom level.  This MS state shows non-negligible coupling to Fock vectors with two atoms in the excited level, which are its shadows.  (c) Probability amplitudes $|\cilmk|^2$ for the first $(N+1)+6N+21(N-1)$ Fock vectors and eigenstates when $V_0=49.8$, $g\approx 7\times10^{-2}$, and $N=8$. (d) Coefficients for the ground state for this case.  Here, even the ground states shows non-negligible coupling to Fock vectors with two atoms in the excited level, being these its shadows.   \label{fig:sh1}}
\end{figure*}

On the other hand, in the Fock regime, as $g$ is increased shadows of the cat states may appear, as discussed above. The curve for which criterion~(\ref{eq:CI}) is satisfied for  
$\chi_{\mathrm{shadow}}=0.15$ is represented  in Fig.~\ref{fig:Vgplaneb}. In Fig.~\ref{fig:sh1} (a) and (c) the probability amplitudes $|\cilmk|^2$ for two examples showing shadows of MS states are shown. These two examples are represented in Fig.~\ref{fig:Vgplaneb} as points 8 and 9, respectively. Fig.~\ref{fig:sh1}(b), represents the coefficients for the eigenstate with index $k=120$ of the first example. This state is a superposition of the Fock vector  
$
\left|0,8\right\rangle \otimes_{l=1,m}\left|0,0\right\rangle,
$ with Fock index $i=1$
and the Fock vectors
$
\left|0,6\right\rangle
\otimes\left|0,1\right\rangle
\otimes\left|0,0\right\rangle
\otimes\left|0,1\right\rangle,
$
$
\left|0,6\right\rangle
\otimes\left|0,0\right\rangle
\otimes\left|0,2\right\rangle
\otimes\left|0,0\right\rangle,
$
 with indices $i=107$ and $i=135$, respectively. These vectors are also coupled to the Fock vectors with indices $i=9$, $i=134$ and $i=155$, corresponding to the MS of the previous ones and those with the same number of atoms in the other well.  As  $g$ is increased further, the shadows of MS  states are more relevant, being noticeable also for the ground state. In Fig.~\ref{fig:sh1}(d) we represent the coefficients for the ground state of the second example. In this case
 the ground state displays non-negligible coupling to its {\it shadow} Fock states. The ground state is then the superposition of the Fock vector $
\left|4,4\right\rangle \otimes_{l=1,m}\left|0,0\right\rangle,
$
with Fock index $i=5$,  to Fock vectors with indices  $i=111$, $i=130$, $i=139$ and $i=151$, i.e, the vectors
$
\left|4,2\right\rangle
\otimes\left|0,1\right\rangle
\otimes\left|0,0\right\rangle
\otimes\left|0,1\right\rangle,
$
$
\left|4,2\right\rangle
\otimes\left|0,0\right\rangle
\otimes\left|0,2\right\rangle
\otimes\left|0,0\right\rangle,
$
and the similar ones obtained after interchanging the well index.
Notice that if $g$ is increased further crossings involving the ground state will take place.  

\begin{figure}\center
\begin{tabular}{c}
\vspace*{-1.25cm}\tabularnewline
\includegraphics[width=\columnwidth]{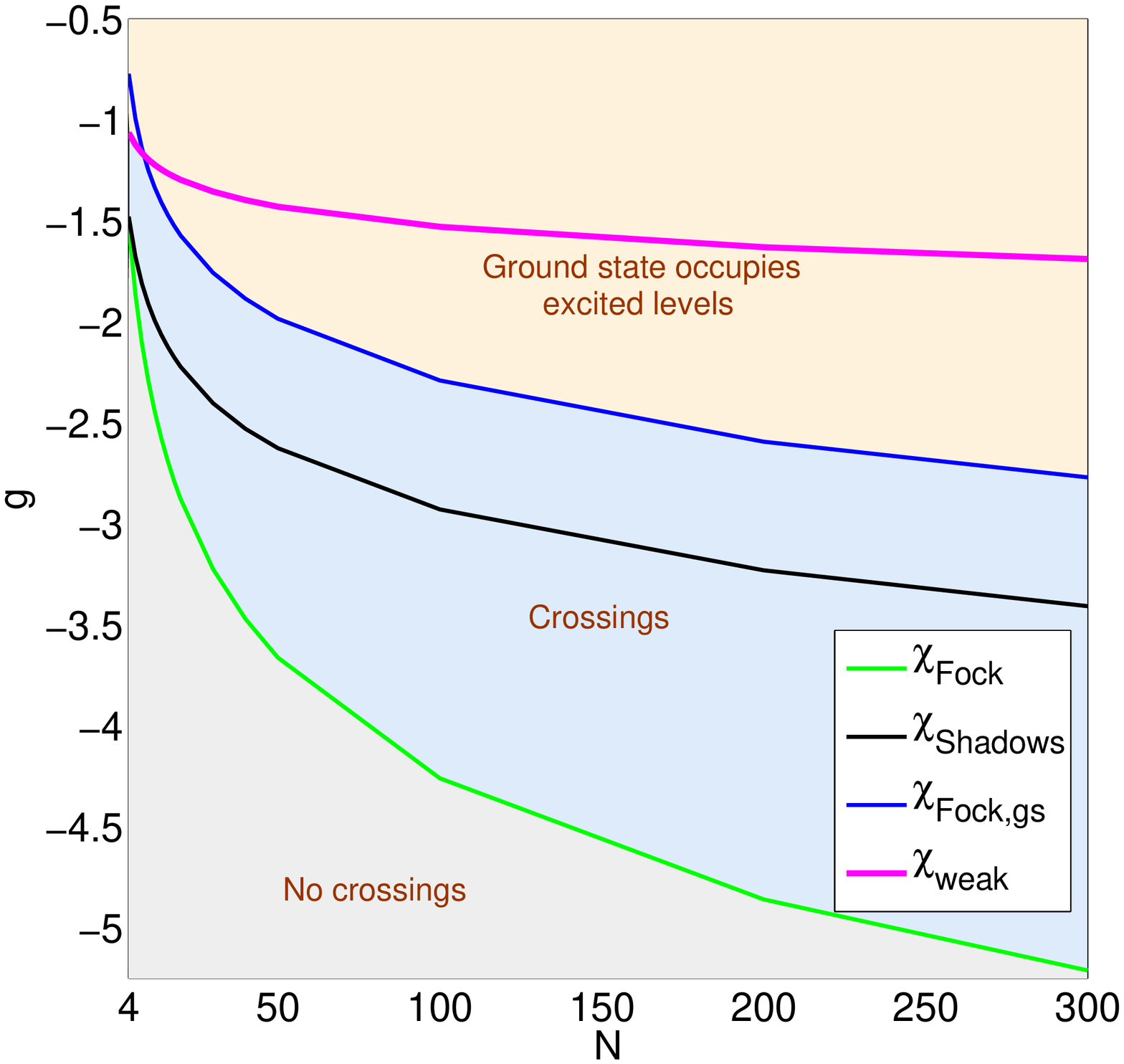}\tabularnewline
\vspace*{-8.75cm}\tabularnewline
(a)\tabularnewline
\vspace*{6.5cm}\tabularnewline
\includegraphics[width=\columnwidth]{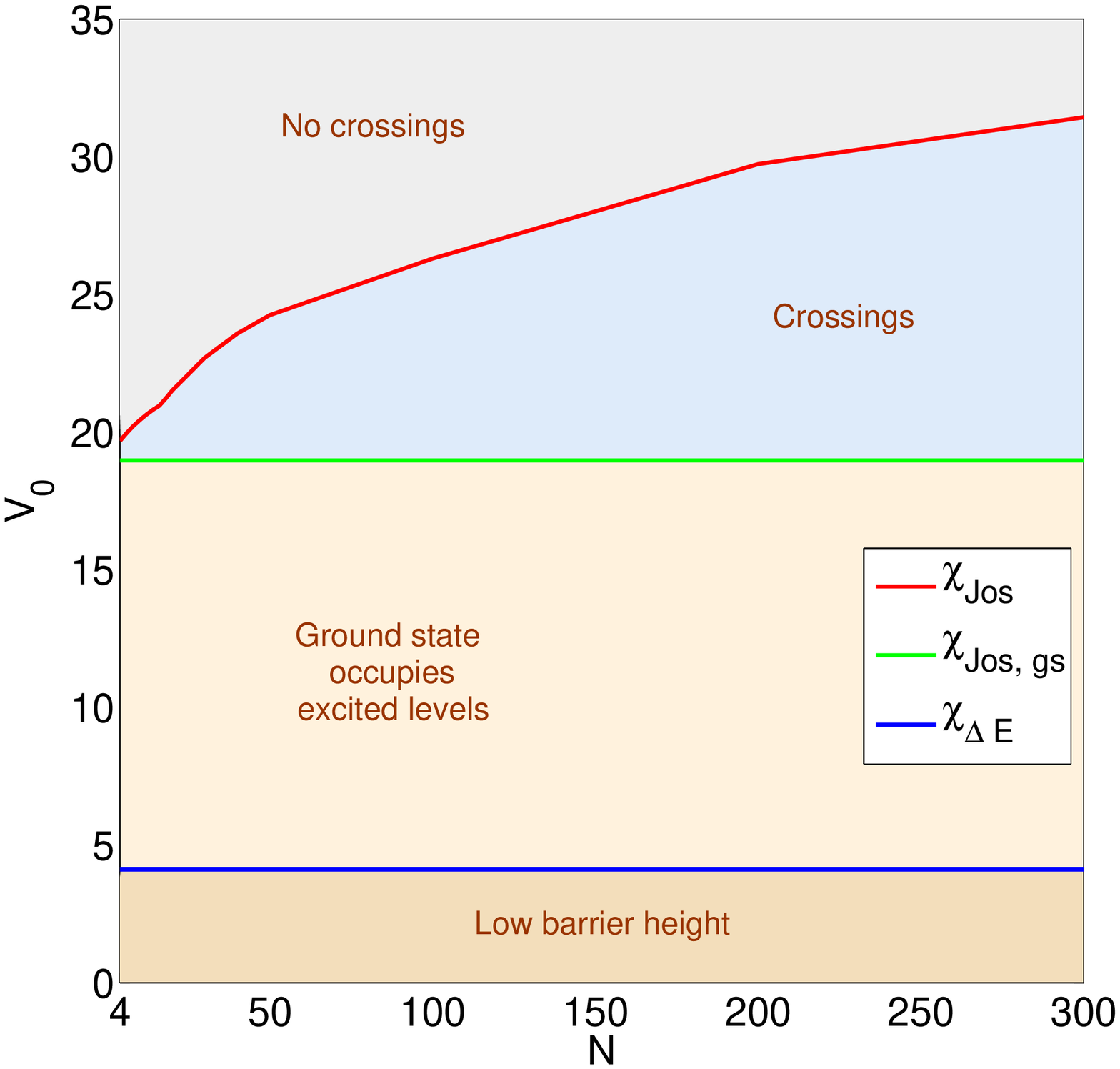}\tabularnewline
\vspace*{-8.75cm}\tabularnewline
(b)\tabularnewline
\vspace*{7.25cm}\tabularnewline
\end{tabular}
  \caption{(Color online) {\it Change of the criteria in the high and low barrier limit with $N$. } (a) Coupling constant $g$ (dimensionless) given by criteria (\ref{eq:CE}), (\ref{eq:CF}), (\ref{eq:CI}), and (\ref{eq:CK}) for a big value of $V_0$,  as a function of $N$. (b) Barrier height $V_0$ (dimensionless) given by criteria (\ref{eq:CC}), (\ref{eq:CD}), and (\ref{eq:CM}) for a small value of $g$,  as a function of $N$. In both cases, the boundaries  between different regimes are displaced accordingly for increasing  $N$. \label{fig:MS_N}  }
\end{figure}

Finally, let us show how this scenario changes as $N$ is increased. In the Fock regime, the curves that represent criteria~(\ref{eq:CE}), (\ref{eq:CF}), (\ref{eq:CI}), and (\ref{eq:CK}), each tend to a straight line as $V_0$ is increased. We have checked numerically that as $N$ is increased these curves move to the left without changing their slope. In Fig.~\ref{fig:MS_N}(a)  we represent how the value of $g$ for $V_0=100$ changes as $N$ is increased.  Then, the area for which  the one level approximation is valid is reduced as $N$ grows, while the region for which shadows of MS states can be observed is increased. Using the harmonic oscillator approximation, it is possible to obtain expressions for the criteria~(\ref{eq:CE}), (\ref{eq:CF}), and (\ref{eq:CK}):
\begin{equation*}
\chi_{\mathrm{Fock}}^{\mathrm{approx}}=\frac{1}{p(N^2-1)},\,\,\,\,\chi_{\mathrm{Fock},\, g}^{\mathrm{approx}}=\frac{1}{p(N-1)},
\end{equation*}
and
\begin{equation*}
\chi_{\mathrm{weak}}^{\mathrm{approx}}=\frac{1}{pN^{1/3}},
\end{equation*}
where $p=\pi^{3/2}V_0^{1/4}/\sqrt{32}$. For criterion~(\ref{eq:CE}) an expression cannot be obtained, but a recurrence relation is obtained. These expressions have been validated numerically, showing good agreement with the numerical curves.

On the other hand,  we have checked numerically that the curve that represent criterion~(\ref{eq:CC}), valid for the Josephson regime, moves upward without changing its slope as $N$ is increased (notice that the other two do not depend on $N$). In Fig.~\ref{fig:MS_N}(b)  we represent how the value of $V_0$ for $g=10^{-7}$ changes as $N$ is increased. Once more, the area for which the one level approximation is valid is reduced as $N$ grows. 

\section{Summary and Discussion}
\label{Sec:conclusion}

We developed a Fock space picture of the stationary states of a system of ultracold bosons in a 3D double-well potential using a two-level, eight-mode approximation. These modes are 3D single particle eigenfunctions with on-well angular momentum  $\ell$ and $z$-component of the angular momentum $m$.  
We have identified all the processes relevant in such a picture.   First, familiar processes occur, such as interaction of atoms in the same  well and the same or different level, or hopping betwen atoms in the same level and different wells.  On the other hand, other less common processes play a fundamental role. These are the hopping   between pairs of atoms in the bottom level with $m=0$ and atoms in the excited level with $m=0$  and hopping between
pairs of atoms in the bottom or excited level with $m=0$ and one atom in the excited level with $m=1$ and other with $m=-1$. We showed that these hopping processes are related to the interaction and not the hopping coefficients. Therefore, in addition to the level spacing, $\triangle E$, and the hopping and interaction coefficients in the bottom levels, $J_0$ and $ U_{0}$,   other coefficients  have to be considered. These are the interaction coefficients between atoms with the same or different values of $\ell$ and $m$, $ U_{\ell m}^{\ell' m'}$, and hopping coefficient between atoms in different wells, $J_{\ell m}$. 

 We found  that all the coefficients are closely related and, indeed, in the high barrier limit, they can be determined in terms of only three of them, $ U_{0}\equiv U_{00}^{00}$, $J_0\equiv J_{00}$, and $J_1\equiv J_{10}$: see Eqs.~(\ref{Eq:Uscales}) and~(\ref{Eq:Jscales}). Nevertheless, in the general case, the coefficients have to be evaluated numerically. They depend on the particular form of the double well potential and on the coupling constant $\bar{g}$. In this Article, we chose a Duffing potential to illustrate our results numerically,  which allows one to reduce this dependence of the coefficients on the particular geometrical  form of the potential and the coupling constant to only two parameters,  $V_0$ and $g$, defined in Eqs.~(\ref{eq:V0}) and~(\ref{eq:g0}), respectively. $V_0$ is related to the barrier height $\bar{V}_0$ and the distance between wells $z_{\mathrm{min}}$, while   $g$ is related to the coupling constant $\bar{g}$ and $z_{\mathrm{min}}$. 

 BECs in double wells were previously thought to exhibit only two main physical regimes, the Josephson and the Fock regime. These are characterized in terms of the hopping coefficient, the interaction coefficient  and the total number of atoms $N$. We derived new relevant coefficients in the problem, and consequently a number of new regimes were identified. Although many new coefficients are considered, they are determined in terms of only two parameters, allowing one to consider the double well problem in terms of three parameters: $V_0$, $g$, and the number of atoms, $N$.  
For certain values of the coefficients the eigenstates show a HO-like behavior, while for others they are MS states corresponding to the conventional Josephson and Fock regimes, respectively.  These regimes have to be considered separately for the bottom and the excited level.  Moreover, we showed that MS states with non-zero occupation of the excited level can occur. These excited MS states show angular momentum degrees of freedom. Finally, we found a mixed regime, in which the eigenvectors with no occupation of the excited levels are MS states, while the eigenvectors showing solely occupation of the excited levels are HO-like states. We found criteria to distinguish all these regimes, which, for fixed $N$, were represented as areas in the $V_0$-$g$ plane. 

The eigenvectors are also different in another region, in which coupling effects between levels  become important since the interaction energy is comparable to the energy level spacing.  In this regime, the interaction energy is much greater than the tunneling energy and the eigenstates are MS states which mix energy levels, showing shadows of cat-like states.  Lowest order perturbation theory couples states with  atoms solely in the bottom level to states with atoms  in the excited level.  We found the criterion to distinguish this region and represented the corresponding curve in the $V_0$-$g$ plane. 

Moreover, eigenstates involving occupation of the excited level can  emerge among the lowest lying eigenstates, for certain values of the interaction and hopping coefficients. We found the criterion that permits one to identify whether such a crossing takes place for a given set of interaction and hopping coefficients. This criterion permits one to define the region in the plane $V_0$-$g$ for which such a  crossing does not occur.  For certain values of the relevant coefficients, even the ground state of the problem  shows occupation of the excited level. The criterion for such a ground state to exist was found. Again, this determines another region  in the plane $V_0$-$g$ in which this ground state with occupation of the excited level does not occur. 
Consequently, three main regions were identified: the region for which no state with occupation of the excited level emerges among the first $N+1$ eigenstates, the one for which this occurs, and the region for which  even the ground state shows occupation of the excited level. For large $N$, the criterion that determines the first crossing for the excited MS states is approximated as $ N^2U_0/2\triangle E$ while the criterion that determines that the ground states shows occupation of the excited level is approximated as $ NU_0/2\triangle E$. Also, the criterion obtained for the shadows of cat states to be relevant also can be approximated as $ NU_0/2\triangle E$. Then for $ N^2U_0\sim2\triangle E$ a two-level approach is necessary to study excited MS states while for  $ NU_0\sim2\triangle E$ our approximation is inaccurate and alternative treatments become necessary~\cite{Masiello:2005,Streltsov:2006}.  

Finally, we have established how all these criteria change with $N$. As the number of atoms  is increased, the area for which the one-level approximation is valid is reduced. Also the region corresponding to the Josephson regime is reduced,  showing, as expected, that mean field approaches are valid for small interaction $U_0$, provided that $N$ is not big. It may appear counterintuitive that MF approaches to the problem are less valid as $N$ is increased, once $U_0$ is fixed, but one must take into account that the criterion for the  limit of the coherent or Josephson regime is $NU/J \ll 1$, thus involving both variables.  For completeness, the limits of validity of the model have been clearly identified and represented in the same $V_0$-$g$ plane.  

The Fock picture for ultracold bosons in the 3D double well potentials developed in this Article will be used in the future to gain insight in the study of dynamical tunneling ultracold bosons in 3D double wells. Now that all regimes have been clearly identified, one can study the tunneling properties of different initial population imbalances in the different regimes. Then, the possible initial states can show occupation of the excited level and  it is expected that complicated and rich dynamics emerge out of the different regimes.  

\begin{acknowledgments}
We thank Joachim Brand, Ann Hermundstad, and William Reinhardt for useful discussions.  L.D.C. acknowledges support from the National Science Foundation under Grant PHY-0547845 as part of the NSF CAREER program. M.A.G.M acknowledges support by the Fulbright Commission, by Spain's Ministerio de Educaci\'on y Ciencia (MEC), and by the Fundaci\'on Espa\~nola de Ciencia y Tecnolog\'{\i}a (FECYT).
\end{acknowledgments}

\appendix

\section{Eigenfunction of the single particle Hamiltonian}
\label{Sec:SP_eigefunctions}

The eigenfunctions $\tilde{\Phi}(\tilde{\xvec})$ of the single particle Hamiltonian  
\begin{equation}\label{eq:spH2}
 \tilde{H}_{\mathrm{sp}}=-\frac{1}{\pi^{2}}\nabla^{2}+\tilde{V}(\tilde{\xvec}),
\end{equation}
permit one to obtain the coefficients $J_{\ell m}/E_r$ and $U^{\ell m}_{\ell' m'}/E_r$ after the scaling described in Sec.~\ref{Sec:adimensionalization}.  Since the potential is separable we can write $\tilde{\Phi}(\tilde{\xvec})=\tilde{\phi}(\tilde{x})\tilde{\phi}(\tilde{y})\tilde{\phi}(\tilde{z})$, where  $\tilde{\phi}(\tilde{x})$, $\tilde{\phi}(\tilde{y})$, and $\tilde{\phi}(\tilde{z})$, are the eigenfunctions of corresponding one dimensional potential with eigenvalues $\epsilon^{\tilde{x}}$, $\epsilon^{\tilde{y}}$ and $\epsilon^{\tilde{z}}$ respectively.

As discussed in Sec.~\ref{sec:two-level}, for the high barrier limit  $\tilde{\Phi}_{n\ell m}(r,\theta,\varphi)  =   R_{n\ell}(r)Y_{\ell m}(\theta,\varphi)$. For convenience, let us write these functions $\tilde{\Phi}_{\ell m}$ in terms of the 1D functions $\tilde{\phi}(\tilde{x})$, $\tilde{\phi}(\tilde{y})$, and $\tilde{\phi}(\tilde{z})$. The  first two lowest excited 1D eigenfunctions are

\begin{equation}\label{eq:EF0}
\tilde{\phi}_{0}(\tilde{x})=\left(\frac{\alpha}{\pi}\right)^{\frac{1}{4}}e^{-\alpha\tilde{x}^{2}/2}
\end{equation}
and
\begin{equation}\label{eq:EF1}
\tilde{\phi}_{1}(\tilde{x})=\frac{\sqrt{2}\alpha^{\frac{3}{4}}}{\pi^{\frac{1}{4}}}\,\tilde{x}\, e^{-\alpha\tilde{x}^{2}/2},
\end{equation}
where $\alpha\equiv4\pi\sqrt{V_0}$. The corresponding eigenvalues are  $\epsilon_{0}=2\alpha/\pi^2$ and $\epsilon_{1}=4\alpha/\pi^2$. The same eigenfunctions and eigenvalues are valid for the other two coordinates $\tilde{y}$ and $\tilde{z}$. 
Then, the ground state of the single particle Hamiltonian~(\ref{eq:spH2}) is
\begin{eqnarray*}
\tilde{\Phi}_{00}(r,\theta,\varphi)
 & = & \tilde{\phi}_{0}(\tilde{x})\tilde{\phi}_{0}(\tilde{y})\tilde{\phi}_{0}(\tilde{z})= \left(\frac{\alpha}{\pi}\right)^{\frac{3}{4}}  e^{-\alpha r^{2}/2},\\
\end{eqnarray*}
with eigenvalue $E_{0}=3\epsilon_{0}=\frac{6\alpha}{\pi^2}$. Analogously, the first excited eigenfunctions are
\begin{eqnarray*}
\tilde{\Phi}_{10}(r,\theta,\varphi)
 & = & \tilde{\phi}_{0}(\tilde{x})\tilde{\phi}_{0}(\tilde{y})\tilde{\phi}_{1}(\tilde{z})\\
 & = & \sqrt{2}\left(\frac{\alpha}{\pi}\right)^{\frac{3}{4}}\alpha^{\frac{1}{2}}r e^{-\alpha r^{2}/2}\cos(\theta),
\end{eqnarray*}
and
\begin{eqnarray*}
\tilde{\Phi}_{1\pm1}(r,\theta,\varphi)
 & = & \frac{1}{\sqrt{2}}\left(\tilde{\phi}_{1}(\tilde{x})\tilde{\phi}_{0}(\tilde{y})\tilde{\phi}_{0}(\tilde{z})\right.\\
&\pm& \left. i\tilde{\phi}_{0}(\tilde{x})\tilde{\phi}_{1}(\tilde{y})\tilde{\phi}_{0}(\tilde{z})\right)\\
 & = & \mp \left(\frac{\alpha}{\pi}\right)^{\frac{3}{4}}\alpha^{\frac{1}{2}} r e^{-\alpha r^{2}/2} \sin(\theta) e^{\pm i\varphi},
\end{eqnarray*}
with eigenvalue $E_{1}=2\epsilon_0+\epsilon_1=\frac{8\alpha}{\pi^2}$. Notice that we have included in the last expression the Condon-Shotley phase convention, as common in the quantum mechanical literature.
\begin{figure*}
\begin{tabular}{|c|c|c|c|c|}
\hline
(a) $R_{00}Y_{00}$ & (b) $R_{11}Y_{10}$ & (c) $R_{11}Y_{11}$ & (d) $\tilde{\phi}_{0}(\tilde{z})$    & (e) $\tilde{\phi}_{1}(\tilde{z})$
\tabularnewline
\includegraphics[width=3.2cm]{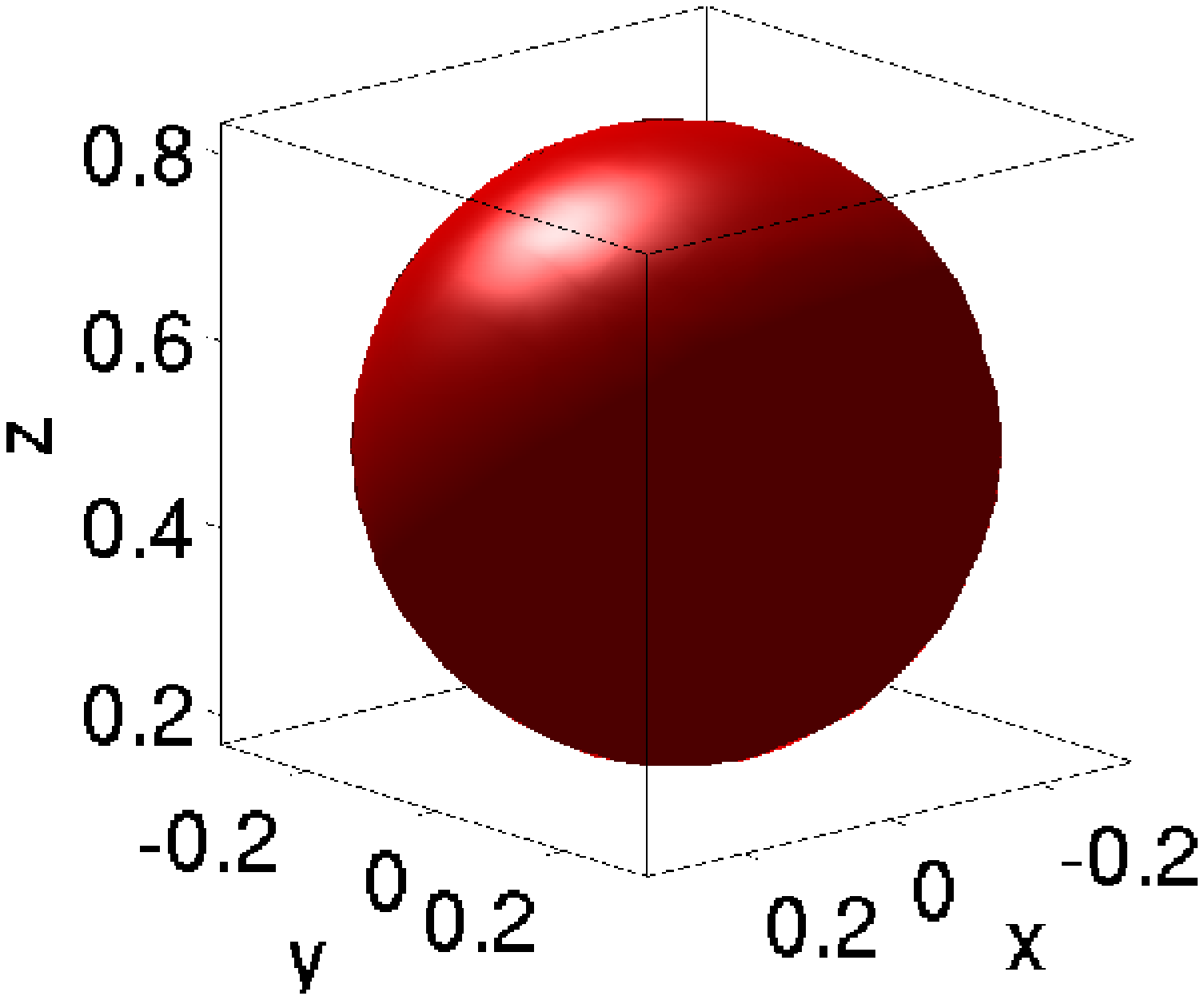}&  
\includegraphics[width=3.2cm]{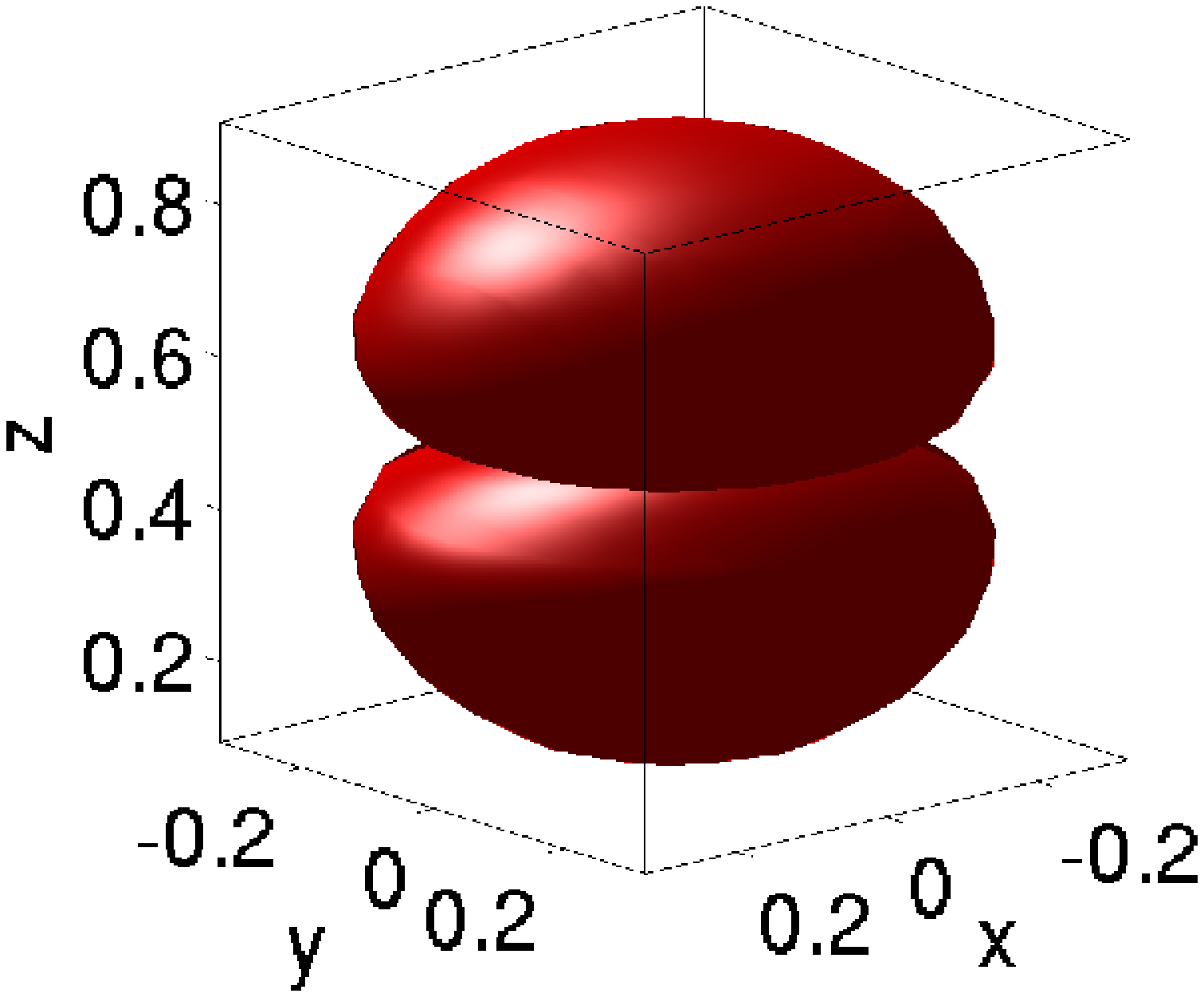}&
\includegraphics[width=3.2cm]{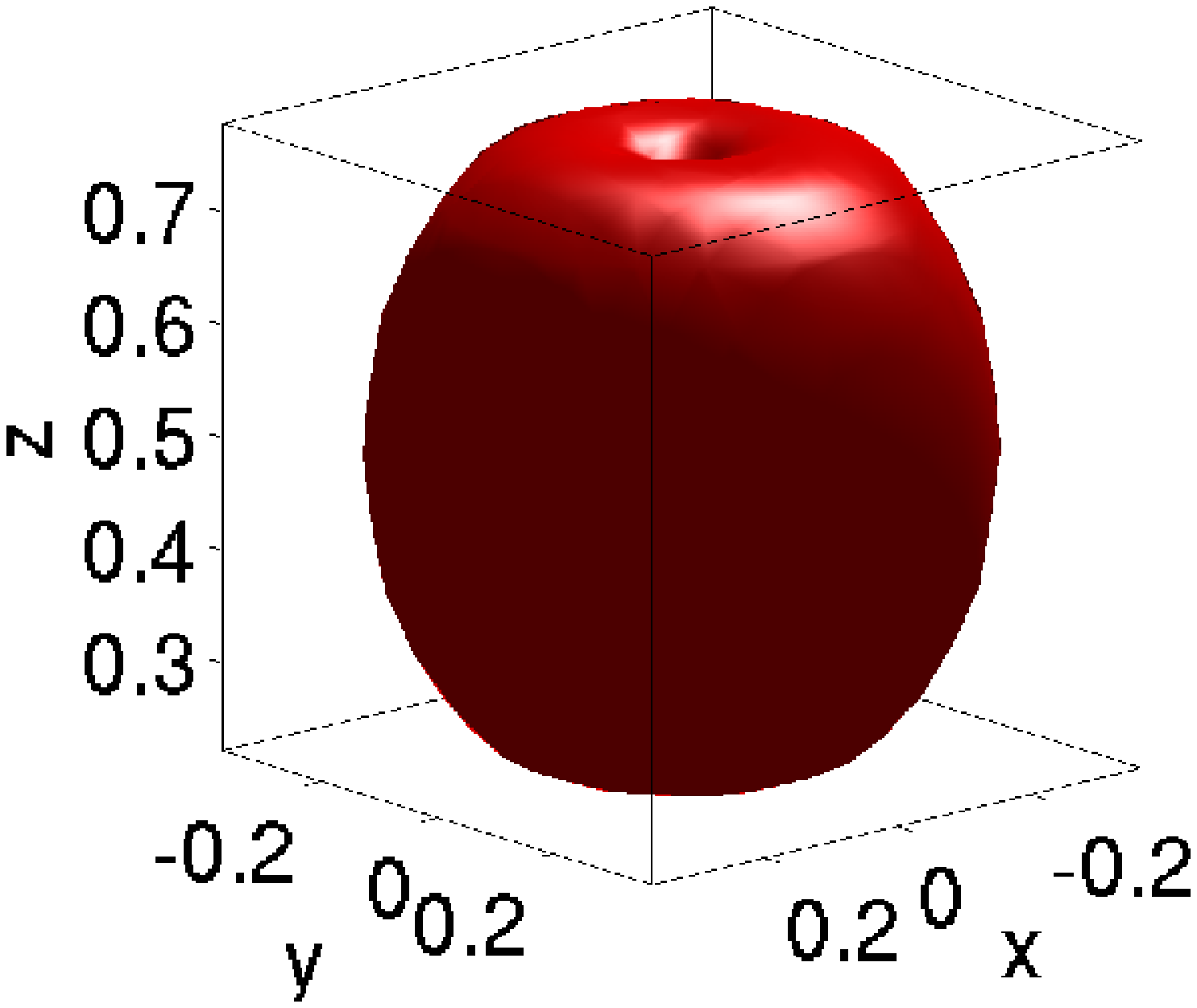}&
 \includegraphics[width=3.5cm]{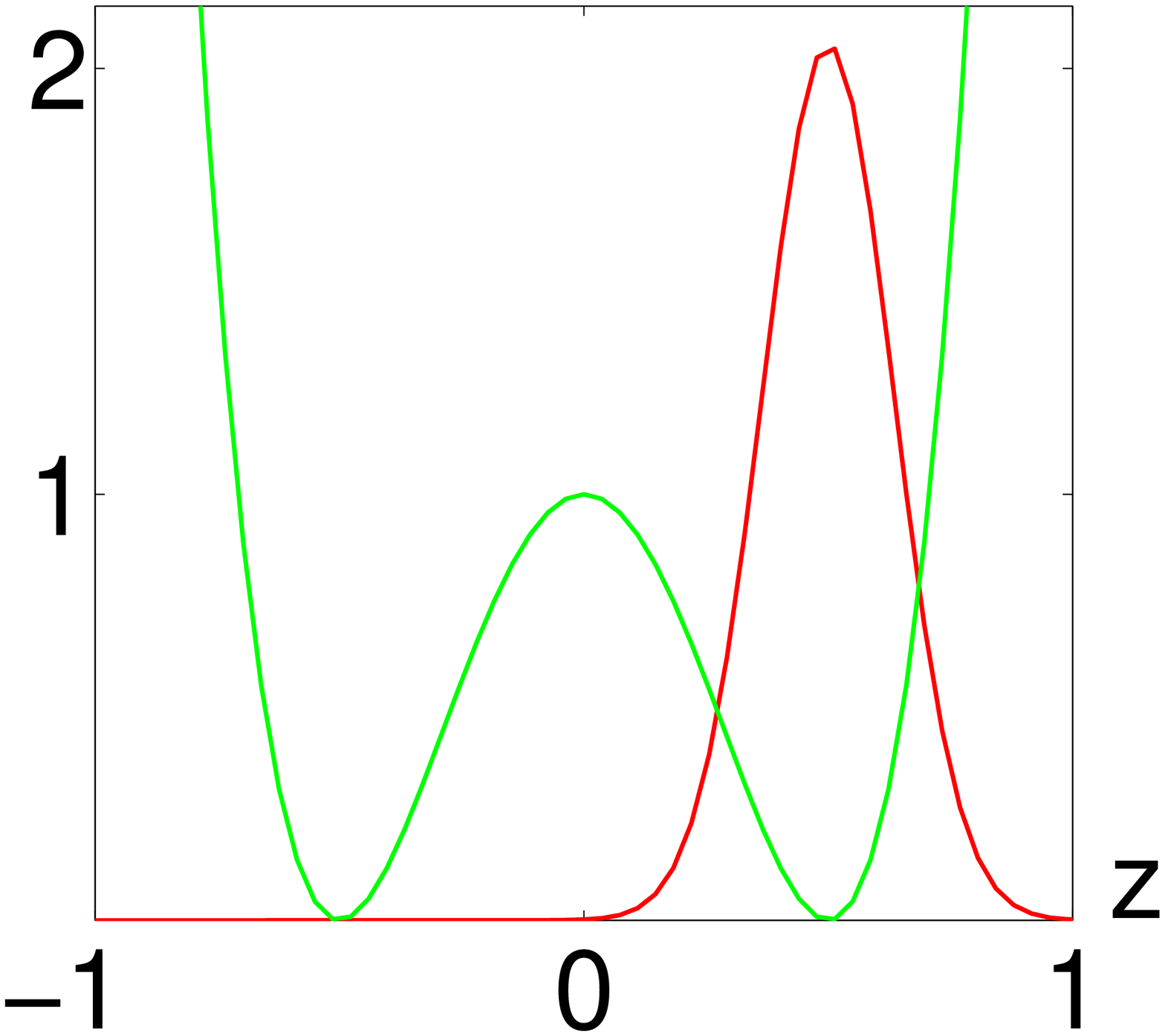}&
\includegraphics[width=3.5cm]{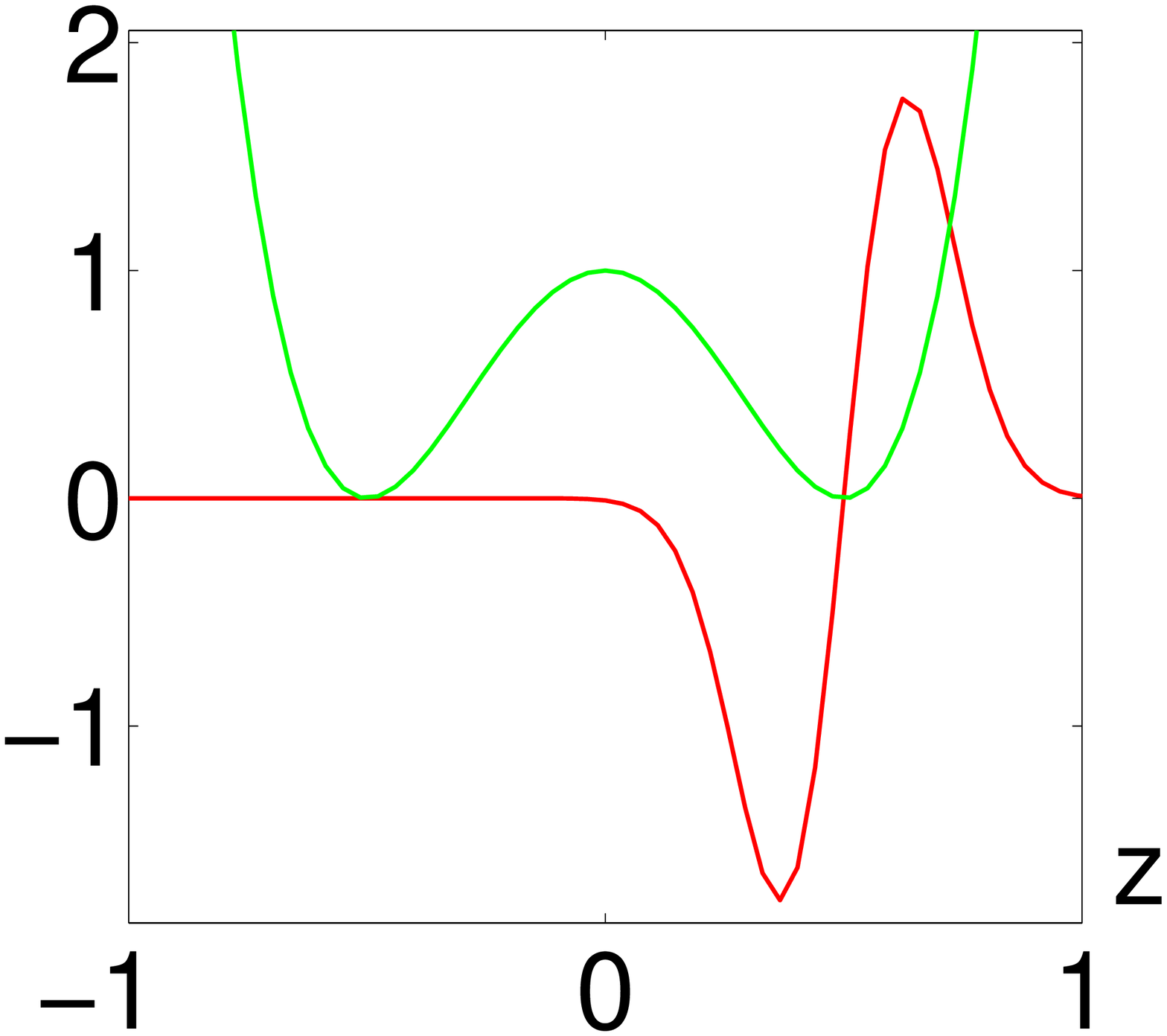}
\tabularnewline
\hline
(f) $\tilde{\Phi}_{00}$ & (g) $\tilde{\Phi}_{10}$ & (h) $\tilde{\Phi}_{11}$ & (i) $\tilde{\phi}^{j=2}_{0}(\tilde{z})$ &(j) $\tilde{\phi}^{j=2}_{1}(\tilde{z})$
\tabularnewline
\includegraphics[width=3.2cm]{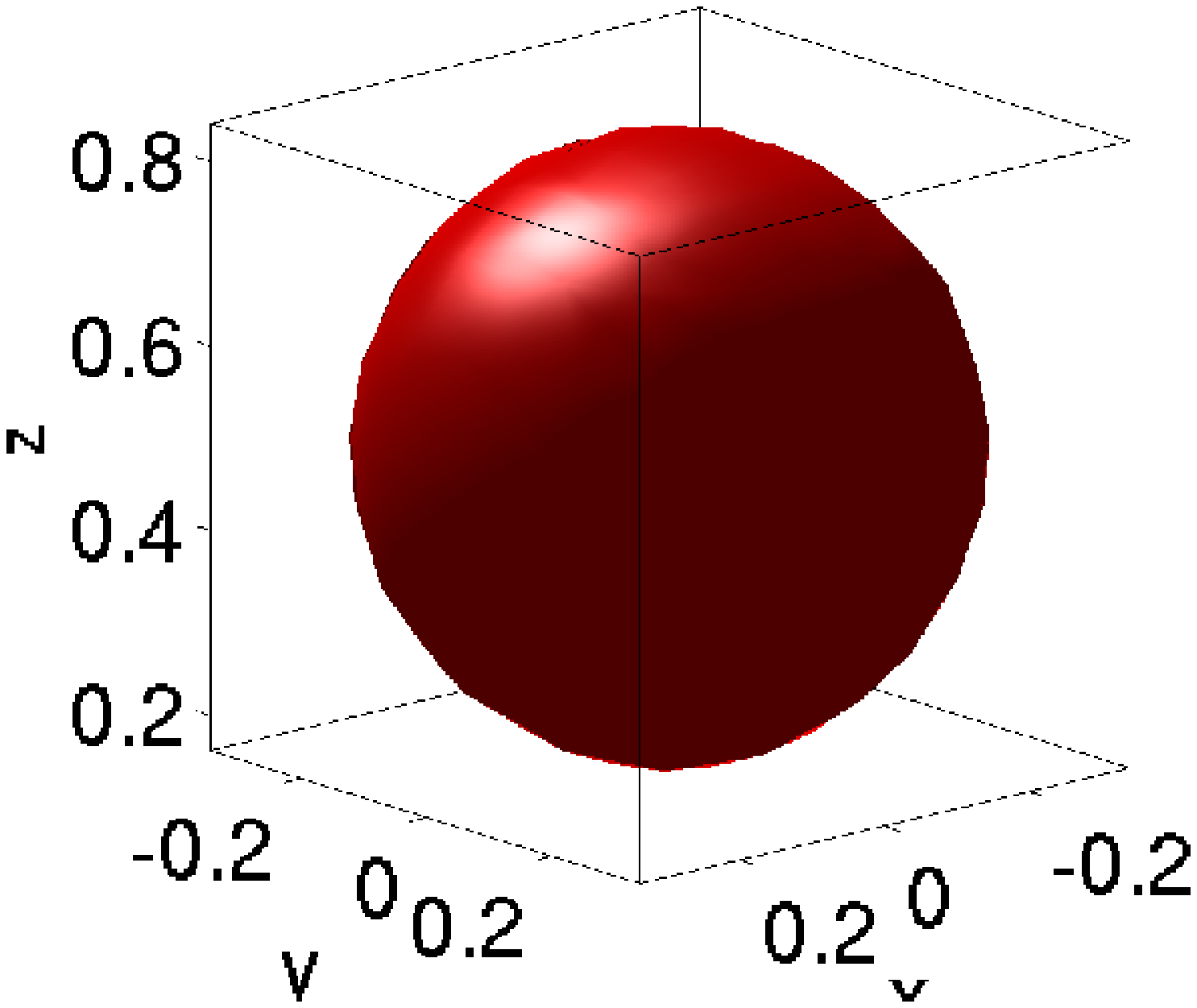}&  
\includegraphics[width=3.2cm]{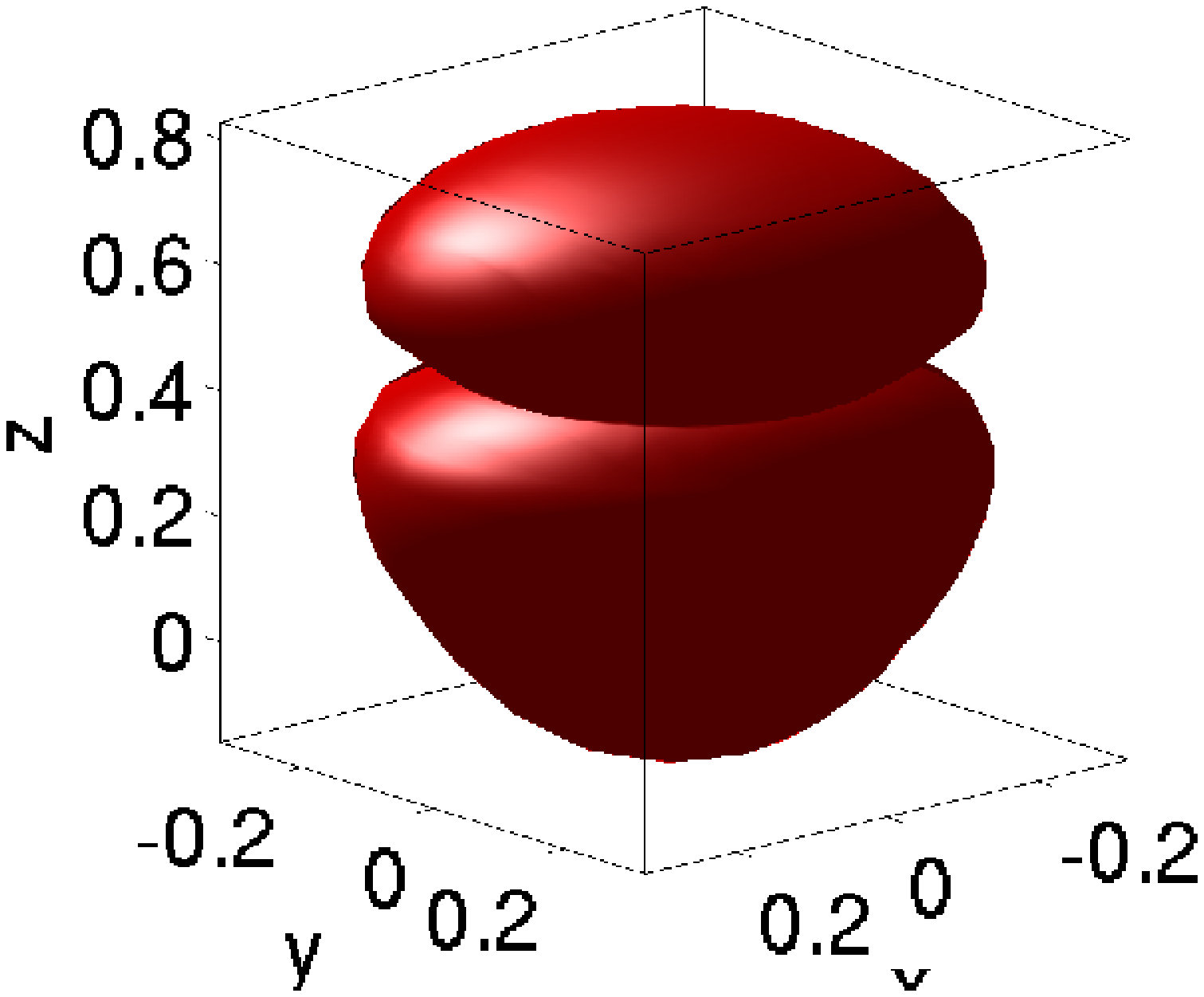}&
\includegraphics[width=3.2cm]{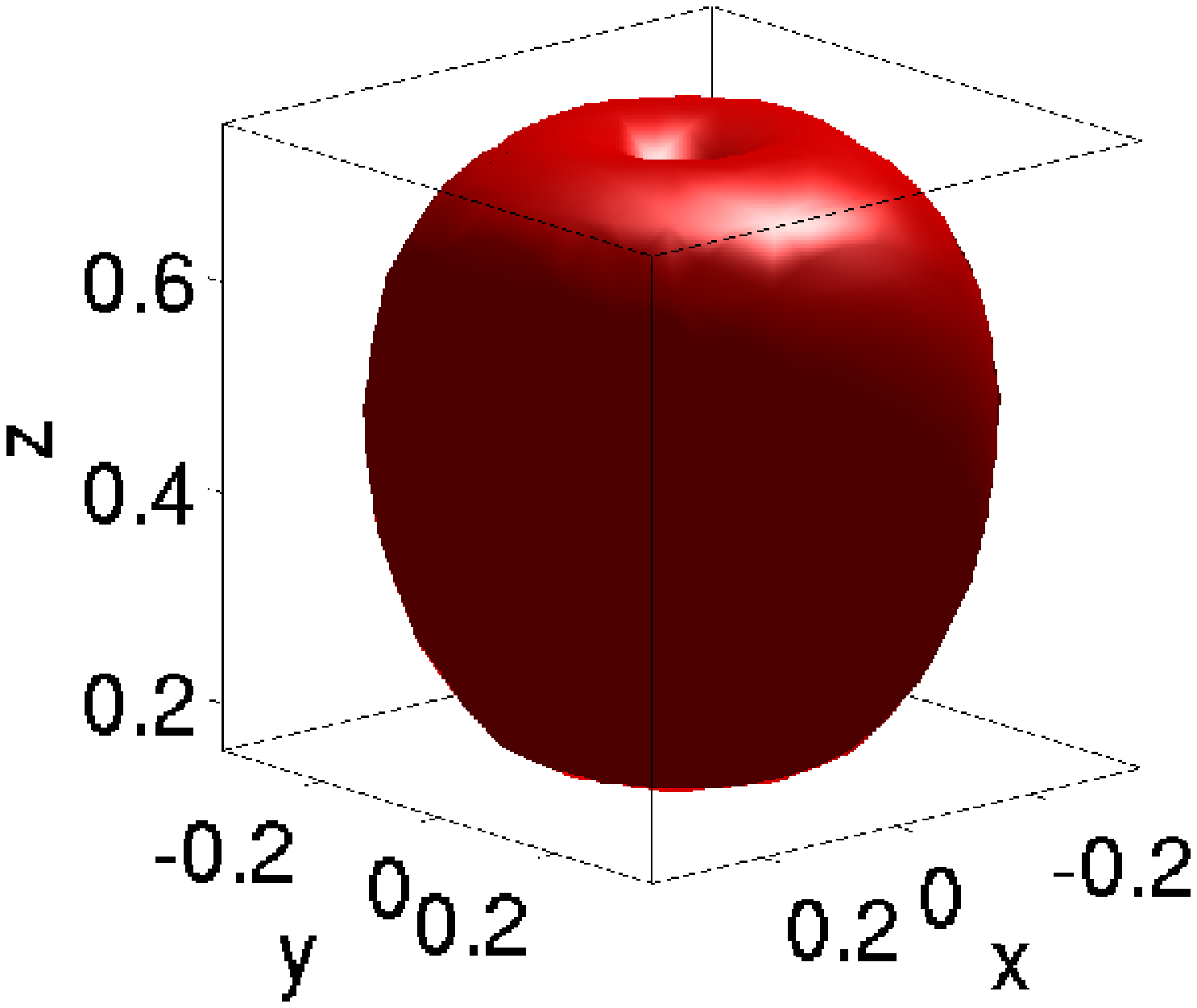}&
\includegraphics[width=3.5cm]{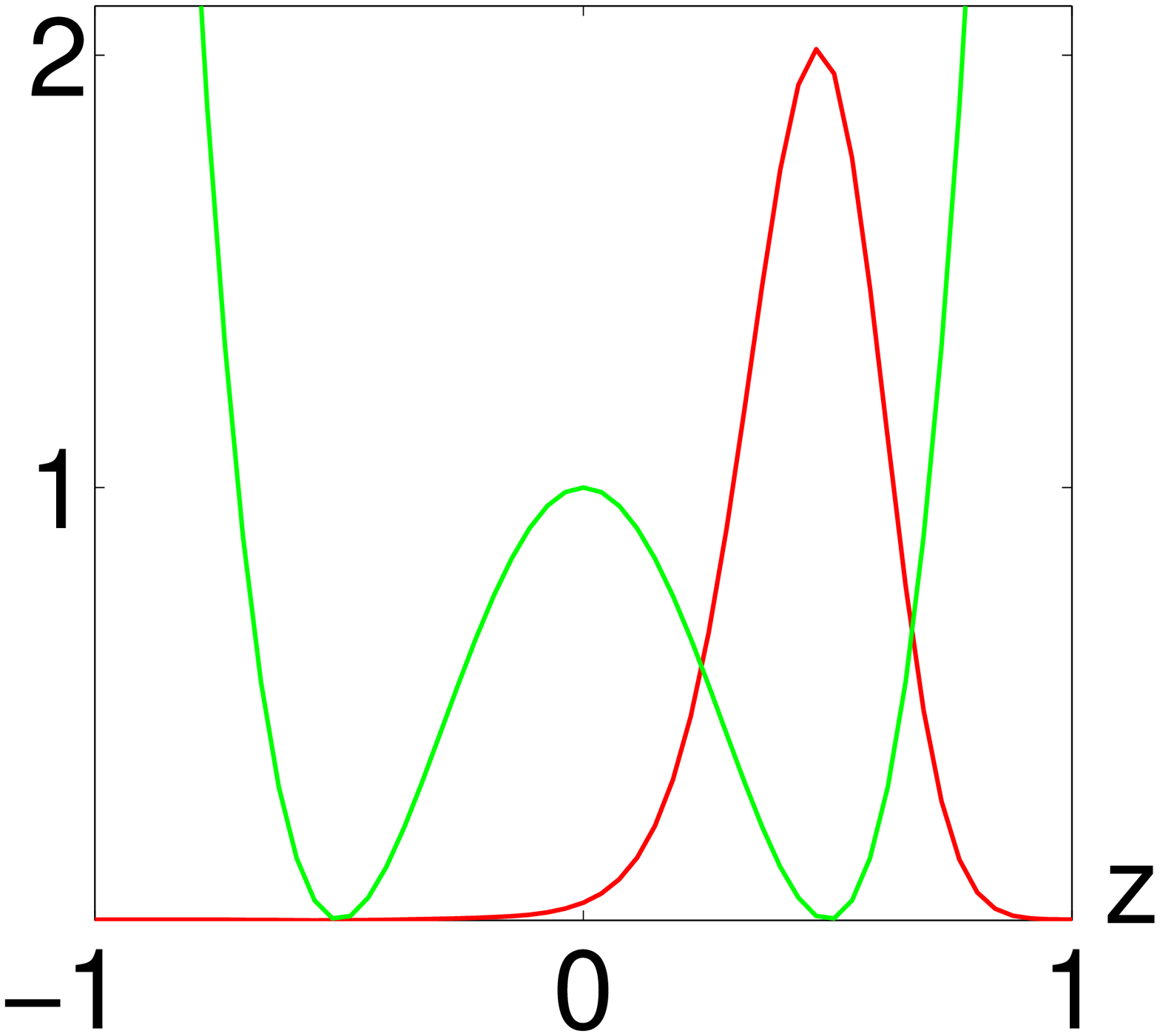}&
\includegraphics[width=3.5cm]{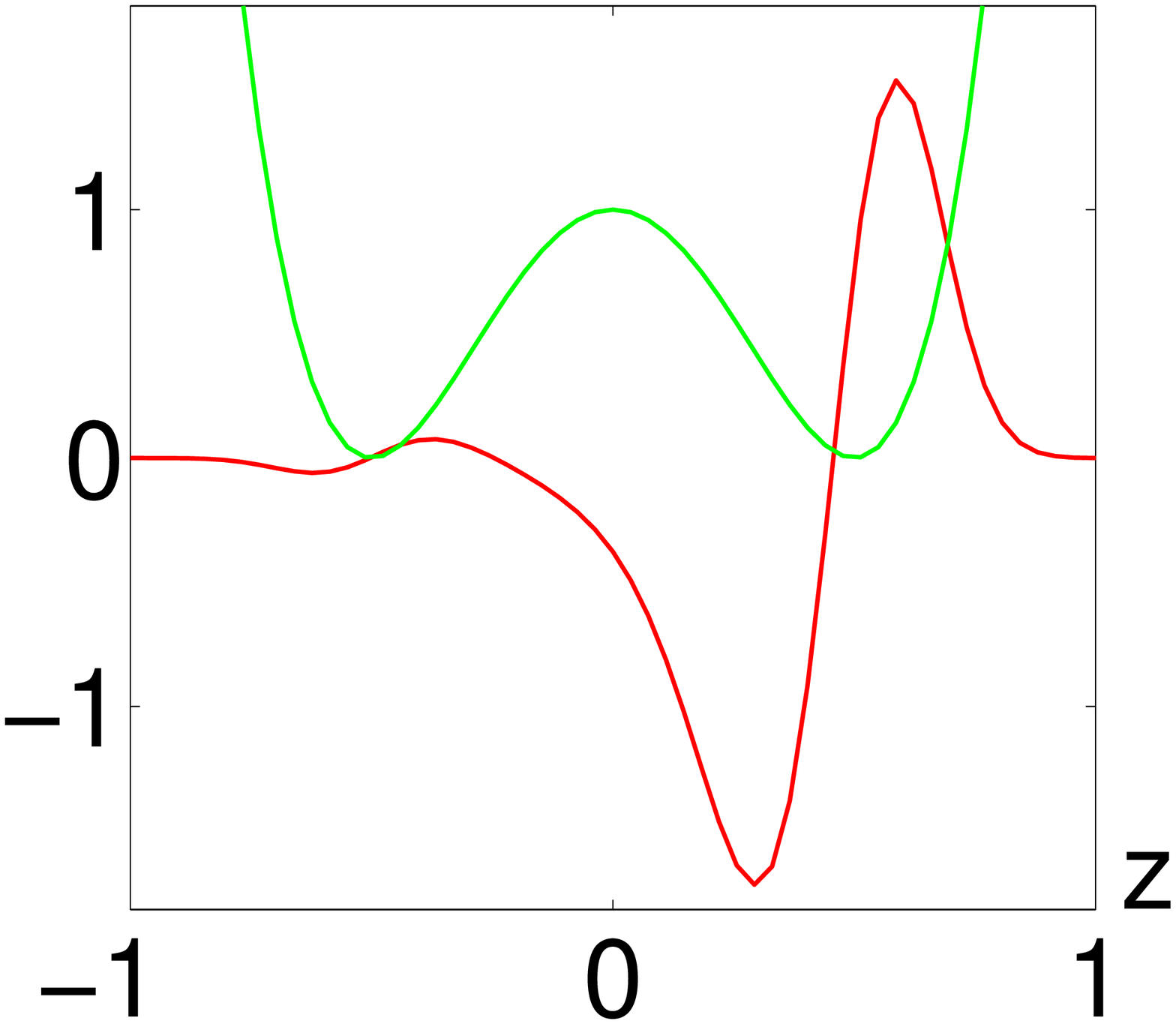}
  \tabularnewline
\hline
\end{tabular}
\caption{(Color online) {\it Eigenfunctions in the high and low barrier limits. }   (a) to (c) represent the analytical functions for the harmonic oscillator approximation while (f) to (h) are the numerically calculated ones for the Duffing potential, all for $V_0=20$. The one-dimensional analytical eigenfunctions are shown in (d) and (e) while the numerical ones are shown in (i) and (j). \label{fig:eigenfunctionstot}}
\end{figure*}

For the low barrier limit this approximation is no longer valid. Nevertheless, we can proceed in the same manner to find the eigenfunctions and eigenvalues numerically. Let us consider the numerical solution localized at well $j$, calculated as  $
\tilde{\phi}^{j}_{0}(\tilde{z})=\frac{1}{\sqrt{2}}\left(\tilde{\psi}^a_{1}(\tilde{z})\pm\tilde{\psi}^a_{2}(\tilde{z})\right),
$
where $\tilde{\psi}^a_{1}(\tilde{z})$  and $\tilde{\psi}^a_{2}(\tilde{z})$ are the first  two numerically calculated eigenfunction of the Duffing potential with eigevalues $\epsilon_1^a$ and $\epsilon_2^a$, respectively. These functions have been calculated with a imaginary time relaxation method. If we consider the plus sign $j=1$ , while $j=2$ in the other case. Similarly 
$
\tilde{\phi}^{j}_{1}(\tilde{z})=\frac{1}{\sqrt{2}}\left(\tilde{\psi}^a_{3}(\tilde{z})\pm\tilde{\psi}^a_{4}(\tilde{z})\right),$ where  $\tilde{\psi}^a_{3}(\tilde{z})$  and $\tilde{\psi}^a_{4}(\tilde{z})$ are  the third and fourth numerically calculated eigenfunctions of the Duffing potential, with eigenvalues $\epsilon_3^a$ and $\epsilon_4^a$, respectively.  We can obtain $\tilde{\Phi}_{jl\pm m}$ using this numerical functions for the 1D functions in the $z$ variable in the expressions given above for the high barrier limit. 
  The corresponding  eigenvalues are $E_{00}=4\alpha/\pi^2+(\epsilon_1^a+\epsilon_2^a)/2$ ,  $E_{10}=4\alpha/\pi^2+(\epsilon_3^a+\epsilon_4^a)/2$, and $E_{1\pm1}=6\alpha/\pi^2+(\epsilon_1^a+\epsilon_2^a)/2$.   Figure~\ref{fig:eigenfunctionstot} shows an example of the numerically calculated  localized eigenfunctions for the low barrier limit using this procedure and the analytical approximation valid for the high barrier limit. As shown, the numerical eigenfunctions  are deformed in the $z$ direction compared to the analytical ones, thus giving higher values of the hopping coefficients $J_{\ell m}$.

\section{Expressions for the coefficients}
\label{Sec:Exprs_Coef}

Let us use the expressions given in App.~\ref{Sec:SP_eigefunctions} to calculate the interaction coefficients. Accordingly, we have
\begin{eqnarray*}
U_{00}^{00} & = & g\int d^{3}\tilde{\xvec}|\tilde{\Phi}_{00}(\tilde{\xvec})|^{4}=gU^{\tilde{x}}_{0}U^{\tilde{y}}_{0}U^{\tilde{z}}_{0},
\end{eqnarray*}
where $U^{\tilde{x}}_{0}=\int d\tilde{x}|\tilde{\phi}_{0}(\tilde{x})|^{4}$, and similiarly for $U^{\tilde{y}}_{0}$ and $U^{\tilde{z}}_{0}$.   In the low barrier limit, we numerically evaluate $\tilde{\phi}^j_{0}(\tilde{z})$. In the high barrier, we  use also the analytical approximation for all the 1D functions and then
\begin{eqnarray*}
U_{00}^{00} & = & g\left(\frac{\alpha}{2\pi}\right)^{\frac{3}{2}}.
\end{eqnarray*}
Analogously, we find
\begin{eqnarray*}
U_{1\pm1}^{1\pm1} & = & g\int d^{3}\tilde{\xvec}|\tilde{\Phi}_{1\pm1}(\tilde{\xvec})|^{2}|\tilde{\Phi}_{1\pm1}(\tilde{\xvec})|^{2}\\
 & = & \frac{g}{4}\left(U^{\tilde{x}}_{1}U^{\tilde{y}}_{0}U^{\tilde{z}}_{0}+U^{\tilde{x}}_{0}U^{\tilde{y}}_{1}U^{\tilde{z}}_{0}+2U^{\tilde{x}}_{01}U^{\tilde{y}}_{01}U^{\tilde{z}}_{0}\right),\end{eqnarray*}
where $U^{\tilde{x}}_{01}=\int d\tilde{x}|\tilde{\phi}_{0}(\tilde{x})|^{2}|\tilde{\phi}_{1}(\tilde{x})|^{2}|$ and $U^{\tilde{x}}_{1}$ is definded as above. Similar expressions hold for $U^{\tilde{y}}_{01}$ and $U^{\tilde{z}}_{01}$.
Finally, we have
\begin{eqnarray*}
U_{10}^{10} & = & g\int d^{3}\tilde{\xvec}|\tilde{\Phi}_{10}(\tilde{\xvec})|^{4}=\tilde{g}U^{\tilde{x}}_{0}U^{\tilde{y}}_{0}U^{\tilde{z}}_{1},\end{eqnarray*}
\begin{eqnarray*}
U_{00}^{10} & = & g\int d^{3}\tilde{\xvec}|\tilde{\Phi}_{00}(\tilde{\xvec})|^{2}|\tilde{\Phi}_{10}(\tilde{\xvec})|^{2}=g\,U^{\tilde{x}}_{0}U^{\tilde{y}}_{0}U^{\tilde{z}}_{01},\end{eqnarray*}
\begin{eqnarray*}
U_{00}^{1\pm1} & = & g\,\int d^{3}\tilde{\xvec}|\tilde{\Phi}_{00}(\tilde{\xvec})|^{2}|\tilde{\Phi}_{1\pm1}(\tilde{\xvec})|^{2}\\
 & = & \frac{g}{2}\left(U^{\tilde{x}}_{01}U^{\tilde{y}}_{0}U^{\tilde{z}}_{0}+U^{\tilde{x}}_{0}U^{\tilde{y}}_{01}U^{\tilde{z}}_{0}\right),\end{eqnarray*}
\begin{eqnarray*}
U_{10}^{1\pm1} & = & g\int d^{3}\tilde{\xvec}|\tilde{\Phi}_{10}(\tilde{\xvec})|^{2}|\tilde{\Phi}_{1\pm1}(\tilde{\xvec})|^{2}\\
 & = & \frac{g}{2}\left(U^{\tilde{x}}_{01}U^{\tilde{y}}_{0}U^{\tilde{z}}_{01}+U^{\tilde{x}}_{0}U^{\tilde{y}}_{01}U^{\tilde{z}}_{01}\right).\end{eqnarray*}
In the high barrier limit, the previous expressions give Eq.~(\ref{Eq:Uscales}). In the low barrier limit, we numerically evaluate $\tilde{\phi}_{\ell}(\tilde{z})$. 

On the other hand, the hopping coefficients are
\begin{eqnarray*}
J_{00} & = & \int d^3\tilde{\xvec}\tilde{\Phi}^{*}_{j00}\left(-\frac{1}{\pi^2}\nabla^{2}+\tilde{V}(\tilde{\xvec})\right)\tilde{\Phi}_{j'00}\\
 & = & E_0\int d^3\tilde{\xvec}\tilde{\Phi}^{*}_{j00}\tilde{\Phi}_{j'00})=E_0J^{\tilde{z}}_{0},
\end{eqnarray*}
where $J^{\tilde{z}}_{0}=\int d\tilde{z}\tilde{\phi}^{j\,*}_{0}\tilde{\phi}^{j'}_{0}$. In the high barrier limit, 
\begin{equation*}
 J_{00} =\frac{6\,e^{-\alpha/4}\left(12+\alpha(\alpha-4)V_0-\frac{(\alpha-2)\alpha^3}{\pi^2}\right)}{\alpha\pi^2}.\label{eq:J0an}
\end{equation*}
Similarly
\begin{eqnarray*}
J_{10} & = & \int d\tilde{\xvec}\tilde{\Phi}^{*}_{j10}\left(-\frac{1}{\pi^2}\nabla^{2}+\tilde{V}(\tilde{\xvec})\right)\tilde{\Phi}_{j'10}=E_{10}J^{\tilde{z}}_{1},\\
\end{eqnarray*}
where   $J^{\tilde{z}}_{1}=\int d\tilde{z}\phi^{j\,*}_{1}\tilde{\phi}^{j'}_{1}$. In the high barrier limit,
\begin{eqnarray*}
 J_{10} & = & \frac{4e^{-\alpha/4}}{\alpha\pi^2}\big \{\alpha\big [36+\alpha(\alpha-6) \big ]V_0 \nonumber\\
& - & \frac{\big [12-\alpha(\alpha-12)\big ]\alpha^3}{\pi^2}-120\big \}
\label{eq:J10an}\end{eqnarray*}
Finally,
\begin{eqnarray*}
J_{1\pm1} & = & \int d\tilde{\xvec}\tilde{\Phi}^{*}_{j1\pm1}\left(-\frac{1}{\pi^2}\nabla^{2}+\tilde{V}(\tilde{\xvec})\right)\tilde{\Phi}_{j'1\pm1}\\
&=&E_{1\pm1}J^{\tilde{z}}_{0}.\\
\end{eqnarray*}
According to the expressions for the eigenvalues given in App.~\ref{Sec:SP_eigefunctions},  in the high barrier limit  $J_{1\pm1}  = 4/3J_{00}$.

\section{Low Barrier Limit}
\label{Sec:expr_HO}

The coefficients $ \cilmk$ defined in Eq.~(\ref{eq:cik}),  are associated with the $i$th eigenstate, with $\Nlm{k} $ atoms in level $\ell$, $z$-component of the angular momentum $m$,  $\nlmi{1} $ atoms in well $j=1$, and $\Nlm{k} - \nlmi{1}$ in well $j=2$. Equation~(\ref{eq:cilmlcoeff}) gives their expression in terms of the  the binomial coefficient $p(\nlmi{1}|\Nlm{k})$ and the normalization constant $a_{\Klmk}(\Nlm{k})$,  which are
\begin{equation*}
  p\left(\!\nlmi{1}\!\left|\Nlm{k}\right.\!\right)= \frac{1}{2^{\Nlm{k}/2}}\sqrt{\frac{\Nlm{k}!}{\nlmi{1}!(\Nlm{k}-\nlmi{1})!}}
\end{equation*}
and
\begin{equation*}
  a_{\Klmk}(\Nlm{k}) = \sqrt{\frac{(\Nlm{k}-\Klmk)!}{\Nlm{k}!\,\Klmk!}},
\end{equation*}
respectively. 

\section{High Barrier Limit}
\label{Sec:HighBarrier}

The unperturbed Hamiltonian is
\begin{eqnarray}
\hat{H}_{U}&=& \sum_{j,m}\!\!\Bigg\{
  \sum_{m'}\!\Big[U^{\ell m}_{\ell m'}
  \nhat{j}{\ell}{m}\left(\nhat{j}{\ell}{m'}\!-\!\delta_{mm'}\right)
  \left(2\!-\!\delta_{mm'}\right)\!\Big]  ,\nonumber\\
&+& E_{\ell}\nhat{j}{\ell}{m}\Bigg\}+\sum_{j,m''}\!\!\Bigg\{ U^{00}_{1m''}4\,U^{00}_{1m''}\,\nhat{j}{0}{0}\,\nhat{j}{1}{m''}\!\Bigg\}.\nonumber\\
\end{eqnarray}
The eigenfunctions of this Hamiltonian
are the Fock basis vectors,
with eigenvalues given by Eq.~(\ref{eq:eigenHB}).
The perturbing Hamiltonian is $\hat{H}_{J}=-\sum_{\ell,m}\hat{H}_{J,\ell m}$, with
\[
\hat{H}_{J,\ell m}=-\sum_{j,m}J_{\ell m}\sum_{j\ne j'}\hat{b}_{j\ell m}^{\dagger}\hat{b}_{j'\ell m}.\]
The dimension of the degenerate subspaces, while depending on the number of atoms in the excited level, is always a multiple of 2. The perturbing Hamiltonian $\hat{H}_{J,\ell m}$ acts on these $2$-dimensional subspaces. Hence, we can diagonalize it in each
subspace. Since
\[
P=\left(\begin{array}{cc}
0 & \left\langle \varphi_{k}|\hat{H}_{J,\ell m}|\varphi_{k'}\right\rangle \\
\left\langle \varphi_{k'}|\hat{H}_{J,\ell m}|\varphi_{k}\right\rangle  & 0\end{array}\right),\]
with
\[
\left\langle \varphi_{k}|\hat{H}_{J,\ell m}^{s}|\varphi_{k'}\right\rangle =0,\]
for $s<N_{\ell m}^{(k)}-2n_{1\ell m}^{(k)}$,  we must use $\left(N_{\ell m}^{(k)}-2n_{1\ell m}^{(k)}\right)$-th
order degenerate theory. In such a case,  we obtain
\[
P'=\left(\begin{array}{cc}
0 & P'_{12}\\
P_{'21} & 0\end{array}\right),\]
with
\[
P'_{12}=\frac{\left\langle n_{\ell m},N_{\ell m}^{(k)}-n_{\ell m}\left|\hat{H}_{J,\ell m}^{N_{\ell m}^{(k)}-2n_{\ell m}}\right|N_{\ell m}^{(k)}-n_{\ell m},n_{\ell m}\right\rangle }{\prod_{n_{1\ell m}=n_{1\ell m}^{(k)}+1}^{N_{\ell m}^{(k)}}\left(\epsilon_{n_{1\ell m}}^{(0)}-\epsilon_{n_{1\ell m}}^{(k)(0)}\right)},\]
and
\[
P'_{12}=\frac{2U_{\ell m}^{\ell m}\left(N_{\ell m}^{(k)}-n_{1\ell m}^{(k)}\right)!\left(\frac{J_{\ell m}}{2U_{\ell m}}\right)^{N_{\ell m}^{(k)}-2n_{1\ell m}^{(k)}}}{n_{1\ell m}^{(k)}!\left[\left(N_{\ell m}^{(k)}-2n_{1\ell m}^{(k)}-1\right)!\right]^{2}}.\]
Therefore, the eigenvectors and eigenvalues of the matrix are
\[
\left(\begin{array}{c}
1\\
\pm1\end{array}\right)\,\,\,\mathrm{and}\,\,\,\pm P_{12}'.\]
The eigenstates include superposition states of the form
\[
\frac{1}{\sqrt{2}}\left(\left|n_{1\ell m}^{(k)},N_{\ell m}^{(k)}-n_{1\ell m}^{(k)}\right\rangle \pm\left|N_{\ell m}^{(k)}-n_{1\ell m}^{(k)},n_{1\ell m}^{(k)}\right\rangle \right),\]
and consequently  symmetric and antisymmetric states appear in nearly degenerate
pairs with energy differences equal to $P'_{12}$. The eigenstate of the complete Hamiltonian is the direct product of these states, with eigenvalue equal  given by Eq.~(\ref{eq:eigenHB}), plus (minus) the energy differences $P'_{12}$ for the symmetric (antisymmetric) state considered.

\section{Interlevel perturbation theory}
\label{Sec:Interlevel_pert}

We consider the unperturbed Hamiltonian as
\begin{eqnarray*}
 \hat{H}_u &=& \sum_{j,l,m}\!\!\Bigg\{
  \!\sum_{m'}\!\Big[U^{\ell m}_{\ell m'}
  \nhat{j}{\ell}{m}\left(\nhat{j}{\ell}{m'}\!-\!\delta_{mm'}\right)
  \left(2\!-\!\delta_{mm'}\right)\!\Big]  \\
&+& E_{\ell}\nhat{j}{\ell}{m}\Bigg\}+\sum_{j,m''}\!\!\Bigg\{
      U^{00}_{1m''}4\,U^{00}_{1m''}\,\nhat{j}{0}{0}\,\nhat{j}{1}{m''}\!\Bigg\},
\end{eqnarray*}
and the perturbing one is then
\begin{eqnarray*}
  \hat{H}_{p} &= &\sum_{j,m}\!\!\Bigg\{(1-\delta_{m0})U^{10}_{1 1}\!\!
  \left[\!\left(\!\bhat{j}{1}{0}{\dagger}\!\right)^2\!
  \bhat{j}{1}{1}{}\bhat{j}{1}{-1}{}
  +\hc\right]\!\!\!\Bigg\}\\
&+& \sum_{j,m''}\!\!\Bigg\{\!
      U^{00}_{1m''}\!\!
     \left[\!\left(\!\bhat{j}{0}{0}{\dagger}\!\right)^2\!
     \bhat{j}{1}{m''}{}\bhat{j}{1}{-m''}{}
     \!+\!\hc\right]\Bigg\}.\\
\end{eqnarray*}
Let us  illustrate the interlevel coupling due to the perturbing Hamiltonian for the $N+1$ eigenvectors with zero occupation of the excited level, i.e.,  states  of the form 
$
  |\phi^{(0)}\rangle =
  |\phi_{00}^{(0)};n_{100}\rangle\bigotimes_{j,1,m}|\phi_{1m}^{(0)};0\rangle.
$ The first order approximation to the ground state $ |\phi^{(1)}\rangle$  gives
\begin{align*}
|\phi^{(1)}\rangle   & =  
c_{1}\left|n_{100}-2,n_{200}\right\rangle \otimes\left|0,0\right\rangle \otimes\left|2,0\right\rangle \otimes\left|0,0\right\rangle\nonumber\\
 & +  c_{2}\left|n_{100},n_{200}-2\right\rangle \otimes\left|0,0\right\rangle \otimes\left|0,2\right\rangle \otimes\left|0,0\right\rangle \nonumber\\
 & +  c_{3}\left|n_{100}-2,n_{200}\right\rangle \otimes\left|1,0\right\rangle \otimes\left|0,0\right\rangle \otimes\left|1,0\right\rangle \nonumber\\
& +   c_{4}\left|n_{100},n_{200}-2\right\rangle \otimes\left|0,1\right\rangle \otimes\left|0,0\right\rangle \otimes\left|0,1\right\rangle\nonumber,\\
\end{align*}
where
\begin{align*}
c_{1} & =  \frac{U_{00}^{10}\sqrt{2\left(n_{100}\right)\left(n_{100}-1\right)}}{U_{00}^{00}\left(6-4n_{100}\right)+2U_{10}^{10}+8U_{10}^{00}+2\triangle E},\\
c_{2} & =  \frac{U_{00}^{10}\sqrt{2\left(n_{200}\right)\left(n_{200}-1\right)}}{U_{00}^{00}\left(6-4n_{200}\right)+2U_{10}^{10}+8U_{10}^{00}+2\triangle E},\\
c_{3} & =  \frac{U_{00}^{1\pm1}\sqrt{\left(n_{100}\right)\left(n_{100}-1\right)}}{U_{00}^{00}\left(6-4n_{100}\right)+8U_{11}^{00}+2\triangle E}, \\
c_{4} & =  \frac{U_{00}^{1\pm1}\sqrt{\left(n_{200}\right)\left(n_{200}-1\right)}}{U_{00}^{00}\left(6-4n_{200}\right)+8U_{11}^{00}+2\triangle E}.\\
\end{align*}
Notice that coefficients $c_3$ and $c_4$ are larger  than the other two. Then, the most relevant modification in the eigenvectors is the coupling to the
vectors $\left|n_{100}-2,n_{200}\right\rangle \otimes\left|1,0\right\rangle \otimes\left|0,0\right\rangle \otimes\left|1,0\right\rangle$ and $\left|n_{100},n_{200}-2\right\rangle \otimes\left|0,1\right\rangle \otimes\left|0,0\right\rangle \otimes\left|0,1\right\rangle$. 
For large $N$ the expressions of these coefficients  become $ NU_{0}/2\triangle E$. Therefore, these coefficients  are negligible insofar as $NU_{0}\ll2\triangle E$. With this perturbing Hamiltonian we have not lifted the degeneracy between the states with the same occupation of opposite wells.  If we  consider also the perturbing Hamiltonian
$
\hat{h}_{J,p}=\sum_{j,\ell,m}\!\!-J_{\ell m}\!\!
  \sum_{j'\neq j}\!\Big[ \bhat{j}{\ell}{m}{\dagger}\bhat{j'}{\ell}{m}{} +\hc \Big]
$, 
the MS states  found in App.~\ref{Sec:HighBarrier}  are obtained. These MS states show coupling to excited states insofar the condition $NU_{0}\ll2\triangle E$ is not satisfied. We call these coupled  states {\it shadows of the MS states}.

\end{document}